\documentclass{iopjournal}
\usepackage{graphicx} 
\usepackage{hyperref}

\usepackage{newtxtext,newtxmath}  
\usepackage{caption}
\usepackage{tikz}
\usetikzlibrary{arrows.meta, positioning}
\usepackage{subcaption}
\usepackage{pgfplots}
\pgfplotsset{compat=1.18}
\usepackage{tikz}
\usetikzlibrary{arrows.meta, intersections, patterns, decorations.pathreplacing}
\usepackage{wrapfig}

\usepackage{amsmath}
\usepackage{bbm}
\usepackage{dsfont}
\usepackage{pgfplots}
\pgfplotsset{compat=newest}
\usepackage{floatrow}
\usepackage[square,numbers]{natbib}
\begin{document}
\fancyhead[L]{}
\fancyhead[R]{}

\title{Transport Novelty Distance: A Distributional Metric for Evaluating Material Generative Models}

\author{
Paul Hagemann$^1$,
Simon Müller$^1$,
Janine George$^{1,2,*}$ and
Philipp Benner$^{1,*}$
}

\affil{$^1$ Bundesanstalt für Materialforschung und -prüfung (BAM)}
\affil{$^2$ Friedrich-Schiller-Universität Jena}\\
\affil{$*$ Correspondence should be addressed to philipp.benner@bam.de and janine.george@bam.de}

\section*{Abstract}
Recent advances in generative machine learning have opened new possibilities for the discovery and design of novel materials. However, as these models become more sophisticated, the need for rigorous and meaningful evaluation metrics has grown. Existing evaluation approaches often fail to capture both the quality and novelty of generated structures, limiting our ability to assess true generative performance.
In this paper, we introduce the Transport Novelty Distance (TNovD) to judge generative models used for materials discovery jointly by the quality and novelty of the generated materials. Based on ideas from Optimal Transport theory, TNovD uses a coupling between the features of the training and generated sets, which is refined into a quality and memorization regime by a threshold. The features are generated from crystal structures using a graph neural network that is trained to distinguish between materials, their augmented counterparts, and differently sized supercells using contrastive learning. 
We evaluate our proposed metric on typical toy experiments relevant for crystal structure prediction, including memorization, noise injection and lattice deformations. Additionally, we validate the TNovD on the MP20 validation set and the WBM substitution dataset, demonstrating that it is capable of detecting both memorized and low-quality material data. We also benchmark the performance of several popular material generative models. While introduced for materials, our TNovD framework is domain-agnostic and can be adapted for other areas, such as images and molecules.

\section*{Introduction}

With the rise of generative AI, there has been a surge in material generative models, which are designed to sample novel and stable materials by adapting to the data distribution on which they are trained (\cite{metni2025generativemodelscrystallinematerials}). However, while there have been many material generative models proposed (e.g. \cite{joshi2025allatomdiffusiontransformersunified, xie2022crystal, MatterGen2025,jiao2024crystalstructurepredictionjoint}), their evaluation is lagging behind. In particular, these works commonly use the so-called SUN metrics (stability, uniqueness, and novelty) to quantify the generated structures individually. This stands in stark contrast to the established evaluation procedure of image generative models, where the Fréchet Inception Distance (FID) \cite{fid} has become the gold standard of evaluation. The FID takes both the training and the generated set, calculates features based on an inception network, and compares the distribution of these features with respect to the Wasserstein metric under a Gaussian assumption. In the molecular world, \cite{fcd} developed the analog Fréchet ChemNet Distance (FCD). An analogue for the Fréchet Distance for materials was developed in \cite{Klipfel_Fregier_Sayede_Bouraoui_2024} based on a GNN, which resembles what we will propose. However, as usual for FID, they operate under Gaussian assumptions. This is not feasible for novelty detection, since a compression of the samples into mean and standard deviation prohibits the detection of memorization.

\begin{figure}[htpb]
    \centering
    \includegraphics[width=1\textwidth]{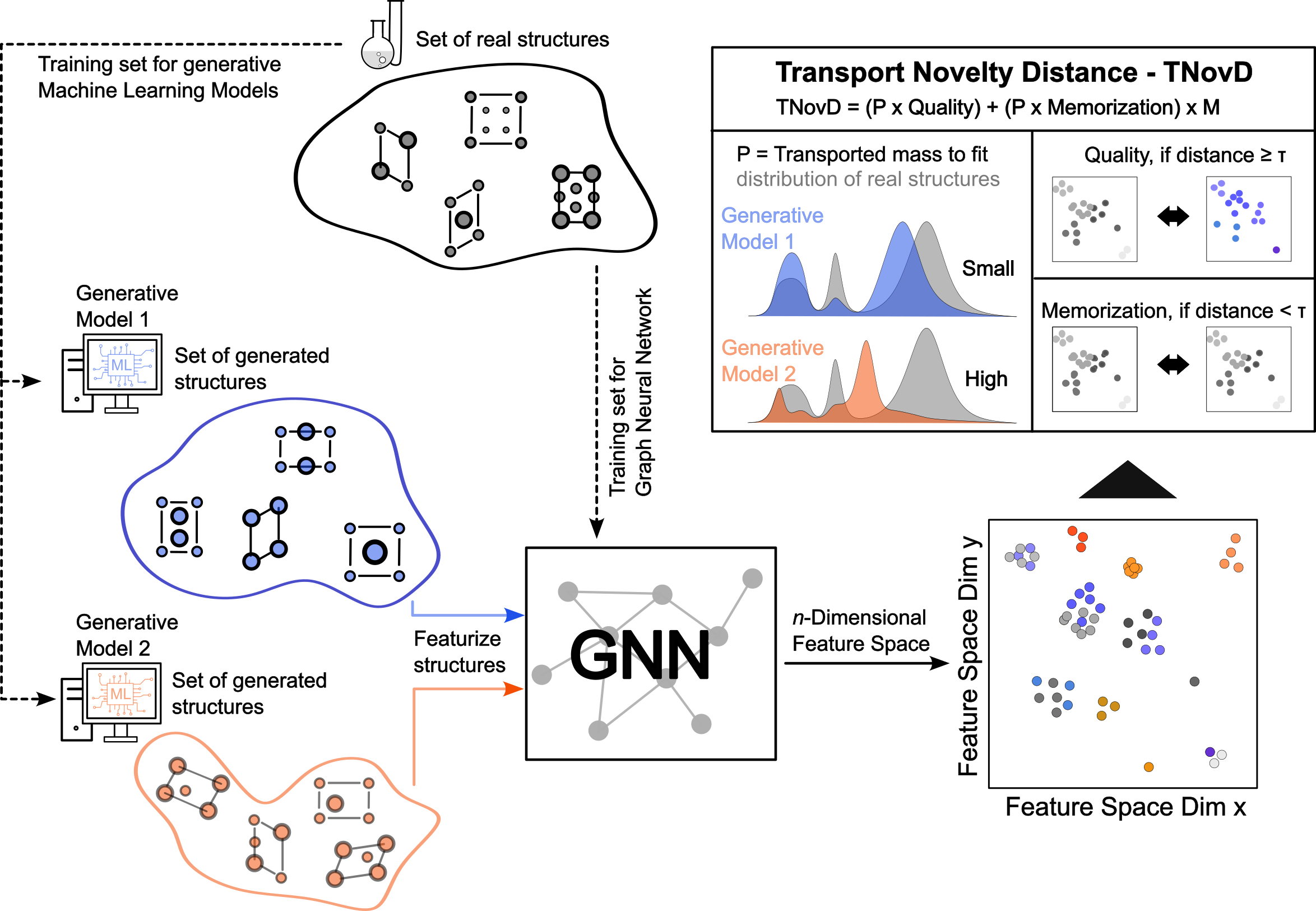}
    \caption{Graphical explanation of the complete workflow used to derive the Transport Novelty Distance between generated sets of materials.}
    \label{fig:tldr}
\end{figure}

In contrast to the FID or the FCD, the SUN metrics do not quantify any distributional properties, and it remains unclear how to combine conflicting signals from the individual indices. Our goal is not to replace these SUN metrics, but rather to take another viewpoint to evaluate material generative models from a probabilistic side, offering a complementary perspective. 

In material discovery, the role of novelty is crucial, but ordered variants of disordered compounds complicate novelty assessments \cite{disordermargraf}.
For the seminal work of MatterGen \cite{MatterGen2025}, there have been recent discussions about whether the material synthesized in this work is part of the training set \cite{disorderdcompounds}.  This example highlights the importance of assessing both quality and novelty simultaneously, preferably within a combined framework.  Similar problems are well-known in the imaging community \cite{Usman_akbar2025}, where, in particular, newer generative models \cite{song2021scorebased} like diffusion models are more prone to memorizing training images, and the FID is unable to capture this. 

In our work, we propose a new metric: the Transport Novelty Distance (TNovD) \footnote{Note that our distance is not a distance in the mathematical sense, but is best when it is small.}. It is based on ideas of optimal transport (OT) \cite{villani} and FID, also resembling ideas from repulsive OT \cite{MarinoGerolinNenna2017}. For its calculation, we take all materials of the training set and the generated set, and embed them in a suitable feature space. This feature space is created by an equivariant graph neural network (GNN) \cite{satorras} that is trained on a classical contrastive learning objective, using the InfoNCE loss \cite{oord2019representationlearningcontrastivepredictive}. This loss nudges the feature representations to be different for different materials, but similar for simple augmentations and supercells of the same crystal structure. Then, we find a coupling between the feature representations of the training set and the generated set. Together with a threshold, this coupling is used to create a customized cost function that penalizes samples that are too close to the training samples, and also penalizes features that are too far away from the training distribution. The complete workflow is illustrated in Fig. \ref{fig:tldr}. Our proposed TNovD is a distributional metric that judges novelty and quality in a holistic way. We also refer to recent advancements in novelty and uniqueness evaluation in \cite{negishi2025continuousuniquenessnoveltymetrics}, where they evaluated these measures in a continuous fashion similar to our OT-based metrics. 

To establish trust in our approach, we performed different sanity checks on the MP20 dataset of the Material Project \cite{materialsproject, xie2022crystal}. We tested the TNovDs ability to detect low-quality materials, which were artificially produced by deformations of the $(x,y,z)$ coordinates, by shearing the lattice and by different kinds of atom substitutions. We also checked for the detection of memorization, rotation, translation and supercells of the training data. We further investigated the novelty/quality performance of the TNovD with the WBM \cite{wbmdataset} dataset, which was produced by iteratively substituting an increasing number of atoms, and then relaxing those structures with density functional theory (DFT), 

Additionally, our TNovD is evaluated on various published material generative models, with samples generated from the MP20 dataset and published by \cite{negishi2025continuousuniquenessnoveltymetrics}. Our work should also be contrasted with \cite{duval2025lematgenbench}, where they compare several models and various metrics capturing distributional properties, such as the maximum mean discrepancy (MMD). 

\paragraph{Related Work}

In the literature, several publications propose new measures of performance for material generative models. In the papers \cite{baird2024, szymanski2025establishingbaselinesgenerativediscovery}, the SUN metrics, as well as coverage, are used for evaluation. This is orthogonal to our contribution, as they perform an \emph{instance-level} rating of the samples, whereas our TNovD is \emph{distributional}. Similarly, in the concurrent \cite{negishi2025continuousuniquenessnoveltymetrics}, continuous novelty and uniqueness scores are provided instead of the usual binary ones. All above mentioned approaches could be combined with our TNovD to gain a more holistic picture of the performance of the models. 

In \cite{ChemEMD}, the chemical earth movers distance (which is a Wasserstein distance) has been introduced, also leveraging ideas from optimal transport to measure distances in terms of composition. It does not take the structural information of the crystal into account. Additionally, it does not yield the distance between \emph{datasets} but rather for two materials. Still, it establishes a precedent for the usefulness of optimal transport in materials sciences. In \cite{onwuli2023}, the use of GNN embeddings is evaluated for assessing the similarity in materials. Closest to our work is a small subset proposed in \cite{Klipfel_Fregier_Sayede_Bouraoui_2024}, where they develop a distributional metric based on GNN embeddings. However, we train our GNN precisely for novelty detection and do not make a Gaussian assumption. Similarly, in \cite{WyckoffDiff}, a Fréchet Distance for capturing symmetries was proposed. However, these two papers are mainly methodological papers that introduce the metric mainly for evaluating their models. 

In the generative modeling community of images, FID is the standard metric, although not undisputed \cite{stein2023}. Memorization is also recognized as a problem \cite{jeon2025understanding} and investigated theoretically \cite{pidstrigach2022scorebased} as well as mitigated practically \cite{jeon2025understanding, zhang2024}. Thereto, kernel methods \cite{zhang2024} or feature log likelihoods \cite{jiralerspong2023feature} are used.

Given the related work, our method aims to fill a gap for a distribution-level metric that combines quality and novelty to evaluate material generative models, complementing many of the existing approaches.

\paragraph{Generative Modeling}
In this paper, we focus on evaluating the quality of generative models designed to create new materials. Independent of their distinct architecture, generative neural networks attempt to model an empirical distribution $P_D$, which is a probability measure on $\mathbb{R}^d$ and is typically specified in the form of samples. Most approaches aim to produce a generating measure $P_G$, which should be close to $P_D$, but not only consist of training samples (see e.g. \cite{dinh2017density} or \cite{song2021scorebased}). Thus, for a given distance measure $D$ between the data distribution and the generating distribution, $D(P_D, P_G)$ should be small but $P_G \neq P_D$. Usually, these models start with some latent distribution $P_Z$ and learn a map $T$ (sometimes implicitly) such that $T_{\#} P_Z \approx P_D$. The latent distribution is thereby a simple distribution, such as the Gaussian. However, it has been shown that the approximation error made by diffusion models leads to "generalization" \cite{pidstrigach2022scorebased}.


\paragraph{Optimal Transport}
The core of our evaluation approach is based on OT \cite{peyre2019}. In general, OT aims to find a coupling between two (abstract) probability measures $\mu$ and $\nu$. It is most instructive to explain OT in the case that both $\mu = \frac{1}{n} \sum_{i=1}^n \delta_{x_i}$ and $\nu = \frac{1}{n} \sum_{i=1}^n \delta_{y_i}$ with $x_i \neq x_j$ for $i \neq j$ and similarly for $y$. Here, $\delta$ is the Dirac measure and these $\mu$ and $\nu$ are \emph{empirical} distributions with masses supported at n points, namely $(x_i)_i$ and $(y_i)_i$. 

Intuitively, OT aims at finding the best permutation between these measures, i.e., a bijective map $\pi$ such that the moved mass is as small as possible.  Let $c = c(x,y)$ denote a generic cost function, commonly the Euclidean or squared Euclidean cost. Mathematically, we can define optimal transport as follows. Denoting by $\Pi^n$ the set of permutations, i.e., the set of bijections from  $\{1,..,n\}$ to $ \{1,..,n\} $, we define the OT map as the permutation

$$\pi_{\text{opt}} = \mathrm{argmin}_{\pi:\Pi^n} \sum_{i}^n c(x_{\pi(i)}, y_i).$$

\begin{wrapfigure}{r}{0.42\textwidth}
    \centering
    \vspace{-5mm} 
    \includegraphics[width=\linewidth]{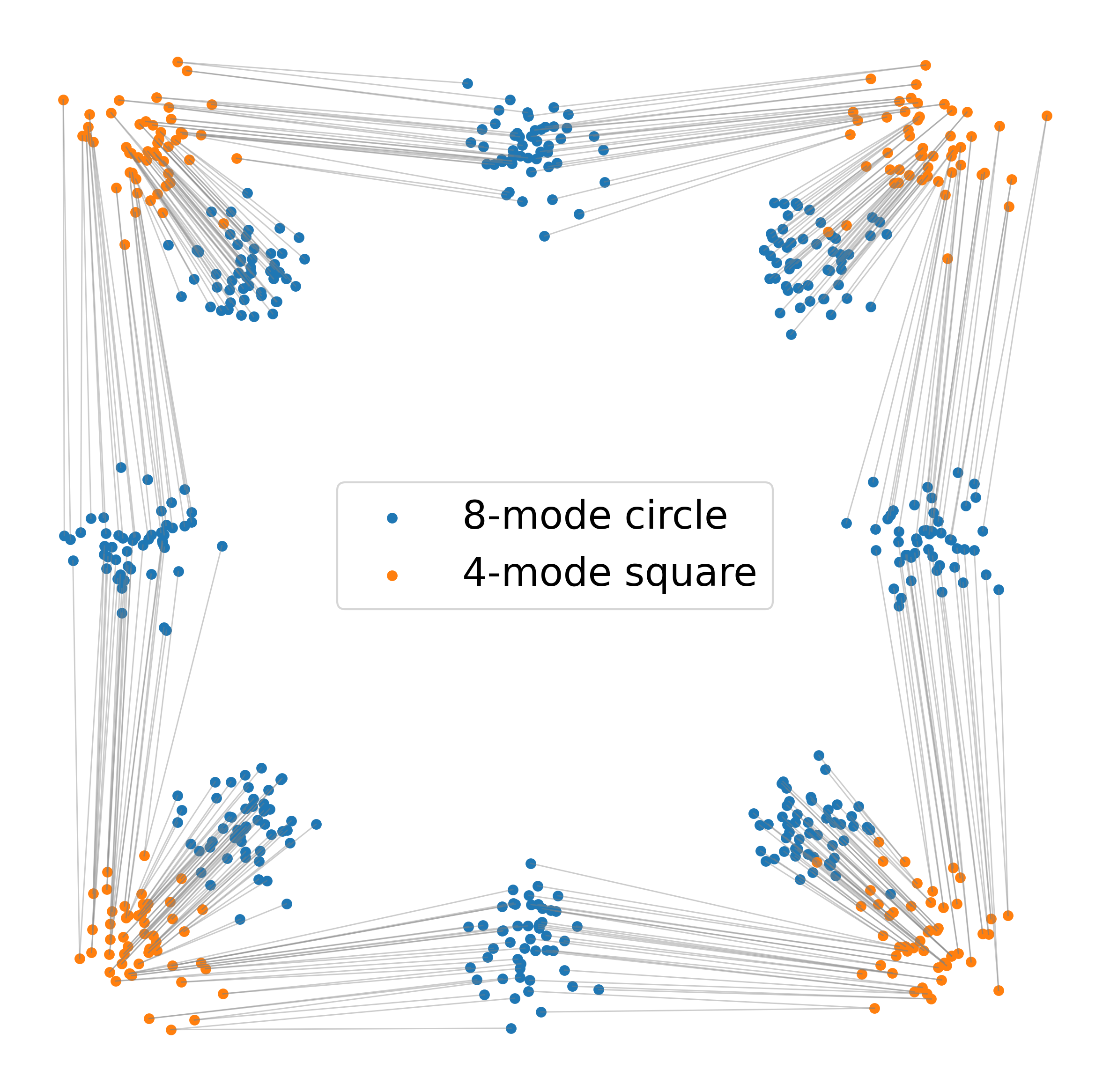}
    \caption{Visualization of an OT coupling between two discrete distributions. The connecting lines indicate the coupling, i.e., the sample assignments.}
    \label{fig:ot_vis}
\end{wrapfigure}

The equation selects the permutation $\pi$ that pairs each $x_{\pi(i)}$ with $y_i$ in such a way that the total matching cost $\sum_{i=1}^n c(x_{\pi(i)}, y_i)$ is minimized. A visualization of such an OT map is shown in Fig. \ref{fig:ot_vis}.

One might ask whether OT is limited to datasets of the same size. Luckily, the framework is strictly more general and also works for continuous data or datasets of different sizes. This is important for our application, where the total number of materials in the training data and in the generated data is rarely identical. In such cases, \emph{couplings} are used to solve the following problem. Assume $\mu = \delta_0$ and $\nu = \frac{1}{2} \delta_{-1} + \frac{1}{2} \delta_1$. Here, the mass needs to be split since "half" of the $\delta_0$, located at 0, needs to move to $-1$ and the other "half" to $1$. However, this could not be modeled with such a matching, and we need a formalism for splitting the masses. This leads to the so-called Kantorovich formulation (see \cite{peyre2019} and \cite{villani} for a more detailed deep-dive). 

A coupling is a probability measure (or distribution) $\pi \in \Pi(\mu,\nu)$ on $\mathbb{R}^{2d}$, where $\Pi(\mu, \nu)$ denotes the set of all probability measures with marginals $\mu$ and $\nu$. More concretely, $\pi \in \Pi(\mu, \nu)$ if $\int_{A \times \mathbb{R}^d} \pi(x,y) dy = \mu(A)$
and $\int_{\mathbb{R}^d \times B } \pi(x,y) dy = \nu(B)$. Now, the optimal transport coupling is given as the minimum over all feasible couplings, i.e., 
$$
    \pi_{\text{opt}} = \mathrm{argmin}_{\pi \in \Pi(\mu, \nu)} \int c(x,y) \mathrm{d}\pi(x,y)\,.
$$
This formulation is the continuous analog of our earlier definition. In essence, we seek a coupling that minimizes the total transportation cost. This implies that a specific $x$ might be coupled with a theoretically more distant $y$ to fit the global minimization perspective, which can be optically observed for individual blue points in Fig. \ref{fig:ot_vis}.

Optimal transport has a deep intrinsic connection to generative modeling. If $\mu$ and $\nu$ possess densities, then the optimal transport problem can be framed as finding the Monge map \cite{villani},
$$
    T_{\text{opt}} = \mathrm{argmin}_{T: T_{\#} \mu = \nu} \int c(x, T(x)) \mathrm{d}\mu(x)\,.
$$
This map $T$ is also a suitable generative model between $\mu$ and $\nu$. In fact, it is a very efficient one, since it needs to move the least mass. This property has often been used in the generative modeling context, e.g. in \cite{tong2024improving, onken2021otflow}.

\paragraph{Graph Neural Networks}
For the developed evaluation metric, the OT is calculated on embeddings (i.e., machine learned representations) of crystal structures, which are created using GNNs. Originally introduced in \cite{Scarselli}, these architectures are now used in many applications and visually introduced in \cite{xie2022crystal, kipf2017semisupervised} and \cite{sanchez-lengeling2021a}.

The main idea of GNNs as message-passing frameworks (\cite{pmlr-v70-gilmer17a}) is based on a graph $G = (V,E)$ with nodes $v \in V$ and edges $(v,w) \in E$, for which we define the following iterations. Our specific SO(3)-invariant GNN follows a simplified version of \cite{satorras} (dropping positional updates). The node features are tuples $(z_i, r_i)$ are given by atom types $z_i$ and positions $r_i$. It proceeds by initializing the node's hidden state $h^0_i = \mathrm{emb}(z_i)$ via an embedding layer of the atoms. 

\begin{align*}
    m^{(t)}_{vw} &= \phi_e \left( h^{(t)}_v \mathbin{\|} h^{(t)}_w \mathbin{\|} \mathrm{RBF}(d_{vw}) \mathbin{\|} d_{vw} \right) && \text{(Edge construction)} \\
    m^{(t)}_v &= \sum_{w \in \mathcal{N}(v)} m^{(t)}_{vw} && \text{(Aggregation)} \\
    h^{(t+1)}_v &= h^{(t)}_v + \phi_h \left( h^{(t)}_v \mathbin{\|} m^{(t)}_v \right) && \text{(Node Update)}
\end{align*}
where $\mathbin{\|}$ denotes concatenation, $\mathcal{N}(v)$ is the neighborhood of $v$, $d_{vw}$ denotes the distance between the positions $r_v$ and $r_w$, $\phi_e, \phi_h$ are neural networks and $\mathrm{RBF}$ denotes radial-basis functions or sinusoidal embeddings, respectively. In each message-passing iteration, edge representations are first computed based on the connected nodes and their features. These edge messages are then aggregated for each target node, and the resulting aggregated message is used to update the node’s information.

We employed contrastive learning to decide whether crystal structures are identical up to certain equivariance or not. More specifically, the InfoNCE loss was used as the loss function of our GNN (\cite{oord2019representationlearningcontrastivepredictive}). This unsupervised technique learns to map positive examples to the same feature and minimize the similarities of negative examples.
The InfoNCE loss for an element/anchor $x$ with positive partner $x^+$ and negative partners $\{x_1,..,x_n\}$ is calculated by minimizing $$ - \log\left(\frac{f(x,x^+)}{ \sum_i  f(x,x_i)}\right).$$
One usually operates in some feature space and chooses $f(x,y) = x^T y$ (for norm-1 vectors), i.e., a simple dot product. If $x$ and $x^+$ are in some learned feature space, then minimizing the InfoNCE loss will increase their similarity and decrease the similarity between $x$ and negative examples. This is often used to learn nice embedding spaces \cite{koker2022graphcontrastivelearningmaterials} or for novelty detection \cite{tack2020csi}. 

\paragraph{Material Parameterization} 
Crystal structures are commonly described based on their unit cell. The unit cell is a repeating unit that can be translated to build the whole crystal structure. The unit cell is described in terms of the lattice and the positions of the atoms within the unit cell. The Crystallographic Information File (CIF) \cite{1991TheCI} is a common way to save and distribute structural information on crystals. CIFs are a text file format in which a unit cell of the crystalline material is saved.  Additionally, information on the crystal structure's symmetry is stored. Thereto, symmetrically inequivalent atoms of the unit cell are stored by name, $(x,y,z)$ coordinates, and site symmetry information. The corresponding lattice, in which the atoms are arranged, is defined and stored by $(a,b,c, \alpha, \beta, \gamma)$, where $a,b,c$ describe the lengths of the lattice and $\alpha, \beta, \gamma$ the angles or only subsets of these parameters, as the others can be deduced from the crystal structure symmetry. We used CIFs parameterization format as input for creating the GNN-based feature space. 

\section*{The Transport Novelty Distance}
\label{sec:TNovD}

In this section, we aim to find a metric that detects whether a generator is producing high-quality (i.e., with statistics similar to the data distribution) and novel samples (i.e., dissimilar from the given training samples). The proposed implementation is based on the packages \cite{pot_package, Ong2012b, pytorchpaper, Fey2019, Fey2025}, the respective code is published in \url{https://github.com/BAMeScience/TransportNoveltyDistance}. 

\paragraph{The generalization-memorization tradeoff}
Our metric should be able to evaluate generative models with respect to both quality and novelty. For this task, generalization is crucial. Philosophically speaking, a generative model generalizes when its induced data distribution $P_G$ matches the (statistical) properties of the \emph{true} data distribution $P_D$. However, we are only provided with a certain number of samples from $P_D$. As a result, we cannot infer the true data distribution and only estimate the likelihood of different generating distributions. 

\begin{figure}[t]
    \centering
    \begin{subfigure}[b]{0.33\textwidth}
        \centering
        \includegraphics[width=\textwidth]{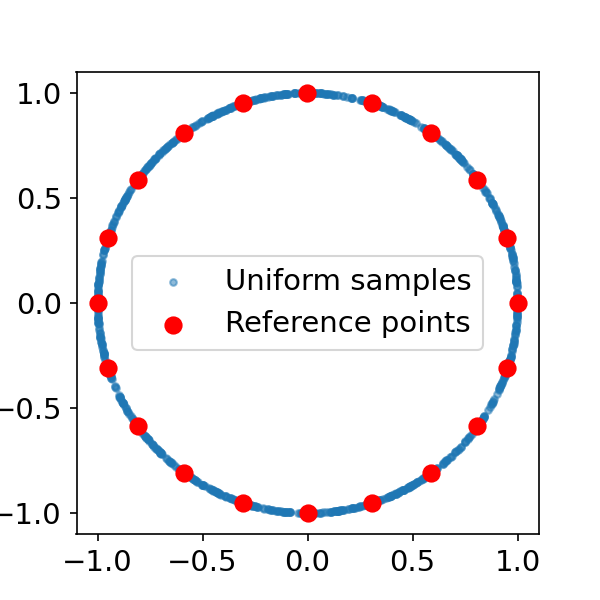}
        \caption{Ground truth}
    \end{subfigure}%
    \begin{subfigure}[b]{0.33\textwidth}
        \centering
        \includegraphics[width=\textwidth]{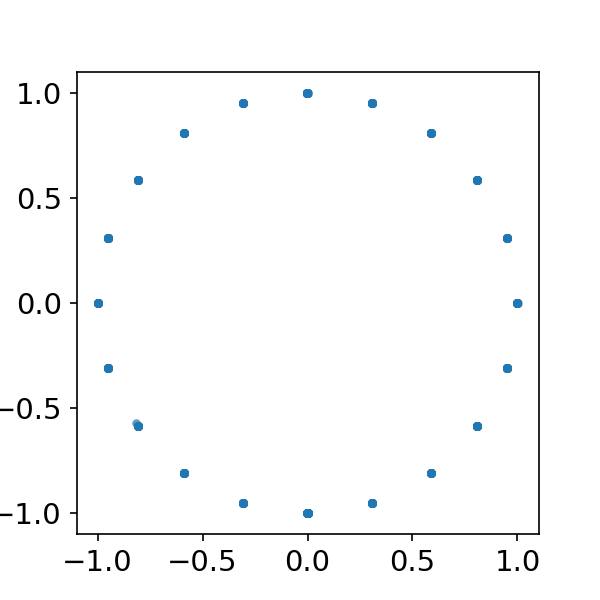}
        \caption{$\sigma=0.00$}
    \end{subfigure}%
    \begin{subfigure}[b]{0.33\textwidth}
        \centering
        \includegraphics[width=\textwidth]{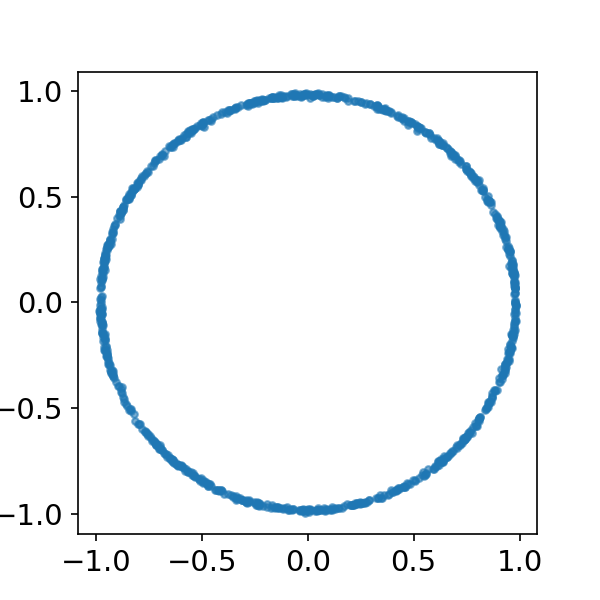}
        \caption{$\sigma=0.17$}
    \end{subfigure}
    \begin{subfigure}[b]{0.33\textwidth}
        \centering
        \includegraphics[width=\textwidth]{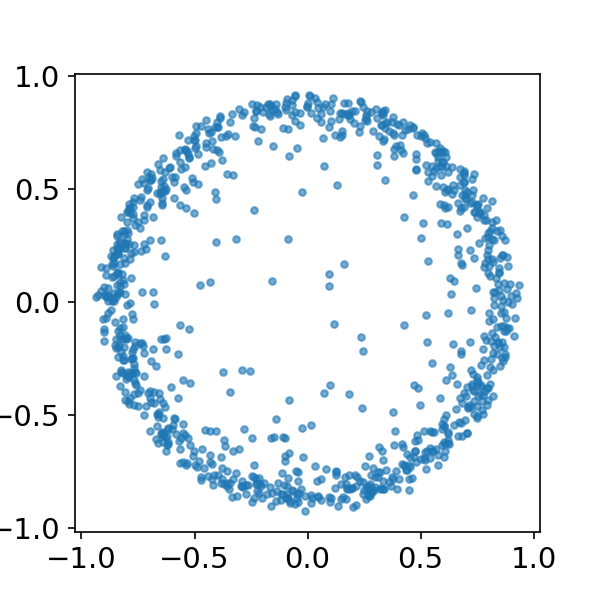}
        \caption{$\sigma=0.40$}
    \end{subfigure}%
    \begin{subfigure}[b]{0.33\textwidth}
        \centering
        \includegraphics[width=\textwidth]{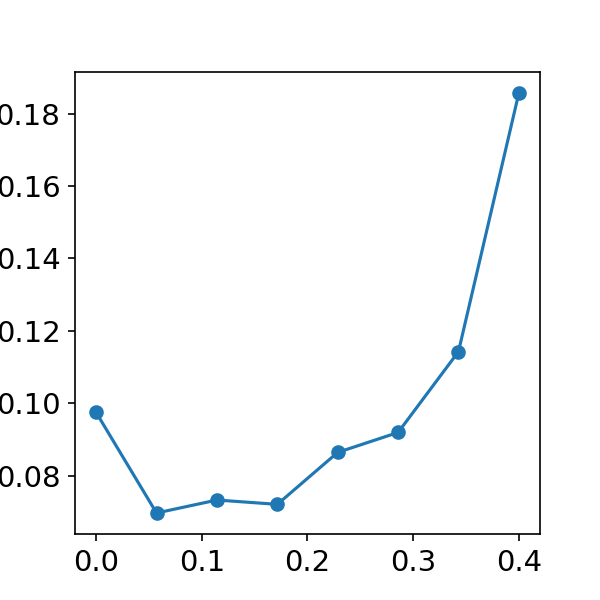}
        \caption{Wasserstein distance}
    \end{subfigure}%
    \begin{subfigure}[b]{0.33\textwidth}
        \centering
        \includegraphics[width=\textwidth]{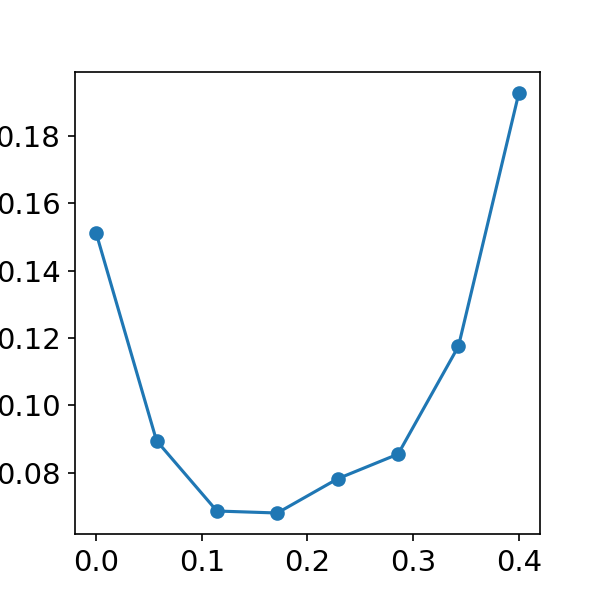}
        \caption{Transport Novelty Distance}
    \end{subfigure}

    \caption{
    (a): True uniform distribution on the circle.
    (b) to (d): Generated samples for increasing smoothing values $\sigma$.
    (e) and (f): Wasserstein distance between validation and generated samples and Transport Novelty Distance, both for varying smoothing values $\sigma$.
    }
    \label{fig:sigma_row}
\end{figure}

We therefore want to introduce a distance measure that only takes a set of training and generated samples and then provides a "distance" that captures whether the model predictions have similar statistics to the training set, without reproducing the training samples directly. For an illustration, we follow ideas from closed-form diffusion/flow matching models presented in \cite{scarvelis2025closedform} and pursue a similar experiment as proposed in \cite{farghly2025diffusionmodelsmanifoldhypothesis}. Here, the true density is a circle on the 2D sphere, from which we sample 2000 points for the training set. A thousand samples of the uniform distribution on the circle are used as the validation set. Then, a closed-form flow matching model is trained for different smoothing parameters $\sigma$, and an identical number of new samples is generated with each model. 
The ground truth density and the flow matching results for different $\sigma$ values are shown in Fig. \ref{fig:sigma_row} (a)-(d). For small smoothing parameters $\sigma$, the model (almost) perfectly reproduces training points. For higher ones, the model generates a more spherical form. 
In Fig. \ref{fig:sigma_row} (e) and (f), the 2D Wasserstein distance and the results of our developed score function are shown. For the former, the generated samples of all different $\sigma$ values were compared to the validation set (Fig. \ref{fig:sigma_row} (e)). For the latter, the generated samples were only compared to the training set (Fig. \ref{fig:sigma_row} (f)).

In both Fig. \ref{fig:sigma_row} (e) and (f), similar U-shaped behavior is observed, with the best value placed in an intermediate regime. For the usual Wasserstein distance, this is in line with observations in \cite{farghly2025diffusionmodelsmanifoldhypothesis, scarvelis2025closedform}.
With well-tuned parameters, our developed score function detects and penalizes memorization of the model close to $\sigma = 0$, resulting in a much larger distance value than seen in Fig. \ref{fig:sigma_row} (e). Similarly, it penalizes unrealistic sample distributions at higher $\sigma$ values. 

\paragraph{Defining the Transport Novelty Distance}
Assume we are given samples from the data distribution, $x_i \sim P_D$, and from the generated distribution, $g_i \sim P_G$. We now define the Transport Novelty Distance (TNovD) to evaluate the quality and novelty of the generated distribution with respect to the data distribution. The TNovD is thereby based on the Wasserstein distance, which is not a distance in the mathematical sense. 

For simplification, we first introduce this score for a given featurizer $F$ and for two sets of samples that both have the same number of elements. In general, this is not necessary. 

\begin{enumerate}
    \item Based on samples $F(x_i)$ and $F(g_i)$, we calculate the optimal transport plan (\cite{peyre2019}) using the Euclidean distance. With the given discrete measures, this amounts to finding an optimal permutation $\pi$ such that the following loss is minimized
    $$\pi^* = \mathrm{argmin}_{\pi:\{1,..n\} \rightarrow \{1,..n\}} \sum_{i} \Vert F(x_i) - F(g_{\pi(i)})\Vert.$$
    \item Then, with this permutation at hand, we calculate the following distance. 
\begin{align*}
    \mathrm{TNovD}(P_D, P_G) = \frac{1}{n} \sum_i \omega\ (\Vert  F(x_i) - F(g_{\pi^*(i)}) \Vert)\,\\ \quad \text{where} \quad 
    \omega(x)
    =
    \begin{cases}
        (\tau - x) M, & x < \tau \ (\mathrm{memorization\ regime)},\\[6pt]
        (x - \tau), & x \ge \tau\ (\mathrm{quality\  regime}).
    \end{cases}\
\end{align*}
\end{enumerate}

One \emph{key detail} is that the coupling in the first step is with respect to standard Euclidean costs, but the second step utilizes this coupling with respect to another cost. This is similar to the strategy in \cite{tameling2021colocalization}. The hyperparameter $\mathrm{M}$ and $\tau$ should be chosen in such a way that the TNovD increases, if bad quality materials or if (almost) replicas of training samples are generated. 

\begin{figure}[h]
    \centering
    \includegraphics[width=0.8\textwidth]{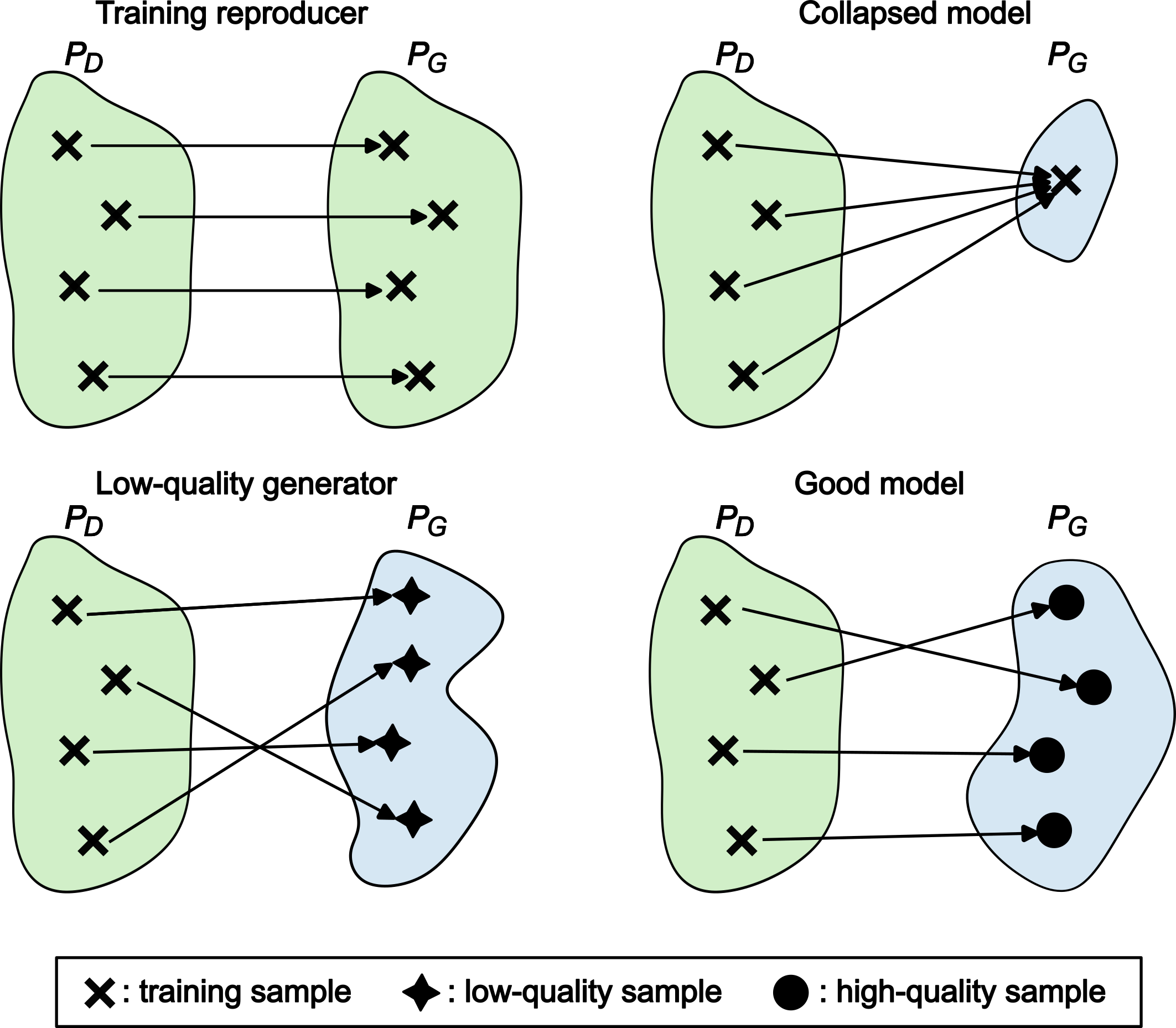}
    \caption{Illustration of optimal-transport matchings between training data $P_D$ (green) and generated samples $P_G$ (blue).  
    Irregular mappings distinguish good vs.~low-quality generations; distorted vs.~aligned shapes indicate poor vs.~faithful model geometry.}
    \label{fig:ot_schematic_final}
\end{figure}

The interaction of the TNovD with specific generative models and datasets is somewhat counterintuitive, which is why we briefly discuss the four most common cases here. A visualization of the four cases can be seen in Fig. \ref{fig:ot_schematic_final}.

\begin{itemize}
    \item The training set: If a model completely reproduces the training set, OT will map every sample to itself. The resulting TNovD will be high due to large values in the memorization regime of the formula. 
    \item A collapsed generator: If a model collapses into a set of the training data, novelty and quality components will both be moderately high, since each generated sample is assigned to one training sample. If multiple generated samples are now mapped to the same training sample, other generated samples must be mapped to training samples much further away, hurting the quality regime of the TNovD. 
    \item A low-quality generator: If a generator is not capable of matching the statistics of the training set, almost none of the samples will fall into the memorization regime, resulting in low memorization. However, values of the quality regime will be high, leading to an overall high TNovD. 
    \item A high-quality generator: a good model will rarely be penalized by the memorization regime, but it can happen. Nevertheless, in most cases, it will generate high-quality samples with features that are close to the training set but not identically memorized. Therefore, values for memorization and quality regime will both be low, resulting in an overall low TNovD. 
\end{itemize}

The distance can be described as deliberately biasing a Wasserstein-type distance so that it actively penalizes reproducing exact training data replications. In contrast to the standard Wasserstein distance, its minimum will not be achieved when we reproduce training data, but in a $\tau$-sphere (in feature space) around the training data. This is an implication of combining novelty and quality into one score: The novelty tries to push the sampled points away from the training data, and the quality tries to attract them. Therefore, the minimizing distribution will lie somewhere between these two extremes. As a result, our TNovD is not suitable as a training objective, since a generator could game it. However, this tradeoff is desirable for any "metric" combining quality and novelty in one, and is precisely what we want for evaluation. This becomes clearer in the limit of "infinite" data: Here, since no point is memorized anymore, $\tau$ goes to $0$. Hence, the usefulness of this TNovD will depend on the choice of $\tau$ and $M$, which we discuss in the section below. 

\paragraph{Heuristic for the Hyperparameter}
Based on the final formula of the TNovD, the hyperparameters $\tau$ and $\mathrm{M}$ are crucial to achieve meaningful results. Here, we present the heuristic used to fit the hyperparameters $\tau$ and $\mathrm{M}$. 

First, we discuss our approach of approximating $\tau$ following some ideas proposed in classical diffusion memorization papers, namely \cite{buchanan2025edgememorizationdiffusionmodels, ye2025provableseparationsmemorizationgeneralization, yoon2023diffusion}). 
They call a point $x$ memorized if $$\Vert x- x^{(1)} \Vert^2 \leq \frac{1}{9} \Vert x-x^{(2)} \Vert^2,$$
where $x^{(1)}, x^{(2)}$ denote the 1- and 2-nearest neighbors from the dataset. After calculating these neighbors for every point in the dataset, we set an estimator $y$ to $1$ if $d_1 \leq d_2/9,$ where $d_1 = \Vert x - x^{(1)} \Vert^2$ and 0 else. Using a grid search, we find the $\tau$ that minimizes the functional $\vert \hat{y}(d_1, \tau) - y(x) \vert$ where $\hat{y}(d_1, \tau)$ is 1 if $d_1 \leq \tau$ and 0 otherwise. This provides us with the best estimator, given the 1-NN distance, for approximating the memorization rule. We remark that this is a purely heuristic rule and leaves a lot of room for both theoretical and practical improvement, but it serves as a good starting point for this paper. 

To find a suitable heuristic for $M$, we make the following formal calculation that is based on having found a good $\tau$ (e.g., with the heuristic above). We assume that for a good $\tau$ and two "good" empirical distributions $P_1, P_2 \approx P_D$ coupled via $\pi$ yield the "distance" random variable (in the feature space) $X_{\text{dist}}$
\begin{align*}
0 \overset{!}{=} \partial_{\tau} \mathrm{TNovD}(P_1,P_2) = M\ P(X_{\text{dist}} \leq \tau) - P (X_{\text{dist}} > \tau).
\end{align*}
Here $P(X_{\text{dist}})$ denotes the probability for the distance measure induced by coupling $P_1$ and $P_2$, i.e., $X_{\text{dist}} = \Vert X^1_{\text{dist}} - X^2_{\text{dist}}\Vert$ for $(X^1_{\text{dist}},X^2_{\text{dist}}) \sim \Pi(P_1, P_2)$. Further, the approach states an "equilibrium" condition for $\tau$. By splitting the training data, two "good" distributions $P_1$ and $P_2$ are created, for which the chosen $\tau$ should be optimal. 
For this optimal state, the derivative of the TNovD$(P_1, P_2)$ with respect to $\tau$ is zero, which results in $M = \tfrac{P (X_{dist} > \tau)}{P(X_{dist} \leq \tau)}.$ Intuitively, this heuristic has desirable properties: The less mass smaller than $\tau$, the larger this constant needs to be to account for this shift. However, this theoretical calculation hinges on the assumption that $M$ should be such that for two datasets coming from the true distribution, the chosen $\tau$ is optimal, meaning that shifting $\tau$ will only increase the score. In our approach, the chosen datasets $P_1$ and $P_2$ are numerically created by splitting the training set into two parts. 

\paragraph{A GNN featurizer}
The proposed TNovD is calculated on abstract features of crystal structures. Thereto, a featurizer $F$ maps all structures from the generated and training sets to a joint feature representation with the purpose of capturing the essential chemical and structural information needed to judge novelty and quality. $F$ should respect important properties, such as being invariant to rotations and translation, as well as agnostic to supercells. 

In our case, the featurizer $F$ is a GNN that is parameterized with $SE(3)$-equivariance and a final pooling head. In contrast to the original architecture of \cite{satorras}, we drop the position updates, making it SE(3)-invariant (similar to \cite{schnet, PhysRevLett.120.145301}). For training the GNN with a feature space size of 32, we downloaded the MP20 dataset with the Train-Validation-Test split employed by \cite{xie2022crystal} and ran ten epochs on all structures from the training set. This took roughly 1.5 hours on an NVIDIA A100 with 80Gb. 

\paragraph{Contrastive Loss} To achieve the expected featurization behavior, we train the GNN using the InfoNCE loss \cite{oord2019representationlearningcontrastivepredictive} with a class of equivariances created from the MP20 train set (\cite{xie2022crystal}). 
Thereto, each batch of training samples is rotated and shifted, which we define as positive examples. 

The remaining samples (i.e., different materials) are defined as negative examples. Mathematically speaking, a batch of samples $\{x_1,...,x_n\}$ is augmented (supercell, rotation, translation) into versions $\{x_1^1,...,x_n^1\}$ and $\{x_1^2,...,x_n^2\}$, resulting in positive pairs $(x_i^1, x_i^2)$. We choose $f(x,y) = x^T y $ as the score function and apply our GNN $F_{\theta}$ with the loss 
$$ L(\theta) = - \sum_{i=1}^n \log \left( \frac{\exp \left( F_{\theta}(x_i^1)^T F_{\theta}(x_i^2) / T \right)}{ \sum_{j \neq i} \exp \left( F_{\theta}(x_j^1)^T F_{\theta}(x_i^1) / T \right) + \sum_{j=1}^n \exp \left( F_{\theta}(x_j^2)^T F_{\theta}(x_i^1) / T \right) } \right). $$
where $T$ is a temperature usually chosen as $0.1$. 

With the defined loss function, training the GNN $F_{\theta}$ pulls the features of positive examples together, while pushing disjoint materials away from one another. In the resulting feature space, physically and chemically similar materials should have similar embeddings.

A severe problem for crystal structure prediction in general is the prediction of supercells with small variations in atomic positions (\cite{criticofgnome}). These predictions can be related to the prediction of different ordered variants of disordered crystal structures. To account for this phenomenon, we added \emph{supercells} as augmentations to our positive examples in the InfoNCE loss. The GNN is therefore trained to map the supercell augments to the same features as the original cell.

\section*{Numerical Results}
A usable score must be robust to common challenges in generative modeling of materials. We therefore tested the developed TNovD on recurring problems in generative machine learning, as well as on artificially altered crystal structures. For the investigated test cases, we treated the validation split of the MP20 dataset from \cite{jiao2024crystalstructurepredictionjoint, xie2022crystal} as newly generated structures, while retaining the MP20 train split as the original training data. We assume that every structure in MP20 is unique and that, therefore, all material in the validation dataset can be considered as novel compared to the training data. The TNovD was calculated for increasingly corrupted MP20 validation data, and the following corruptions were tested. 

\begin{itemize}
    \item \emph{Training Data Reproduction}:
    In principle, generative models would achieve a very good loss when reproducing training data or translations and rotations thereof.  
    To investigate the TNovD for such behavior, we iteratively replaced an increasing amount of the validation data with training data (or rotations and translations thereof), until only training data was involved in calculating the TNovD. For defined replacement ratios, we calculated the novelty score.   
    \item \emph{Noise Pertubations}: Badly trained generative models end up pushing the latent noise into regions where there should be certainty. To test the score in the presence of noisy samples, we added increasing amounts of Gaussian noise to the $(x,y,z)$ coordinates of each atom in the crystal cell, which were then pushed into the respective lattice. 
    \item \emph{Lattice Shear}: To test for another kind of noise, we altered the respective crystal lattice with an iteratively growing scale factor, and checked whether the TNovD can capture this deformation. 
    \item \emph{Random and Group Substitutions}: Additionally, we randomly substituted (without relaxing) atoms with a certain probability by either random atoms or by random atoms belonging to the same group in the periodic table. 
    \item \emph{Supercell probability}: To test the invariance of the TNovD towards supercells, we randomly created a supercell from the validation set structure with a certain probability. The supercell-scale is one of $(1,1,2),(1,2,1),(2,1,1), (2,1,2), (1,2,2), (2,2,1), (2,2,2)$. 
\end{itemize}

This is by no means a complete list, but an easy way to check if the score is able to account for common unrealistic, memorized, or invariant materials. 

\begin{figure}[htpb]

    \centering
    \begin{subfigure}[b]{0.45\textwidth}
        \centering
        \includegraphics[width=\textwidth]{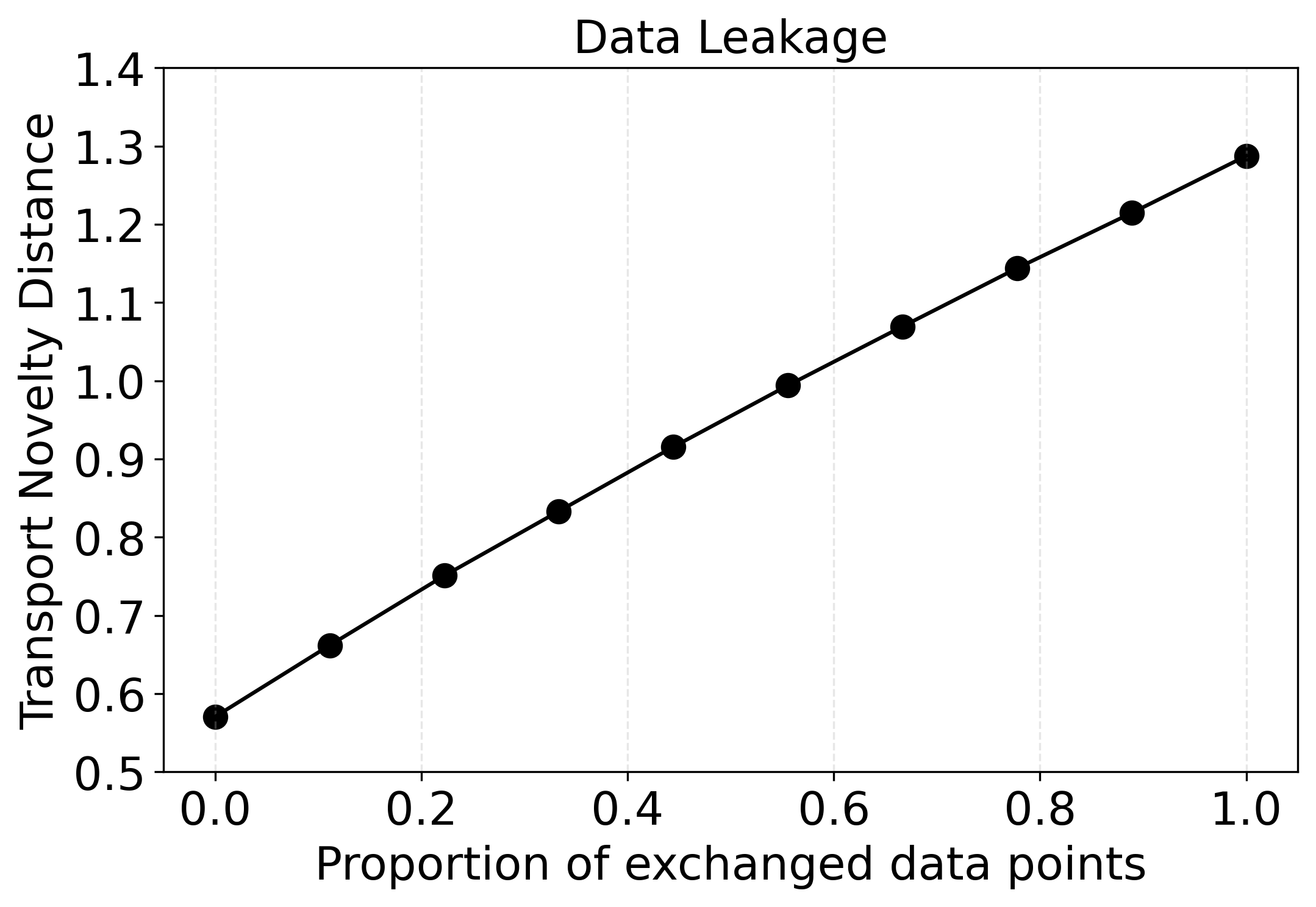}
        \caption{Validation data replaced by training data and its TNovD.}
        \label{fig:data_leakage}
    \end{subfigure}
    \hfill
    \begin{subfigure}[b]{0.45\textwidth}
        \centering
        \includegraphics[width=\textwidth]{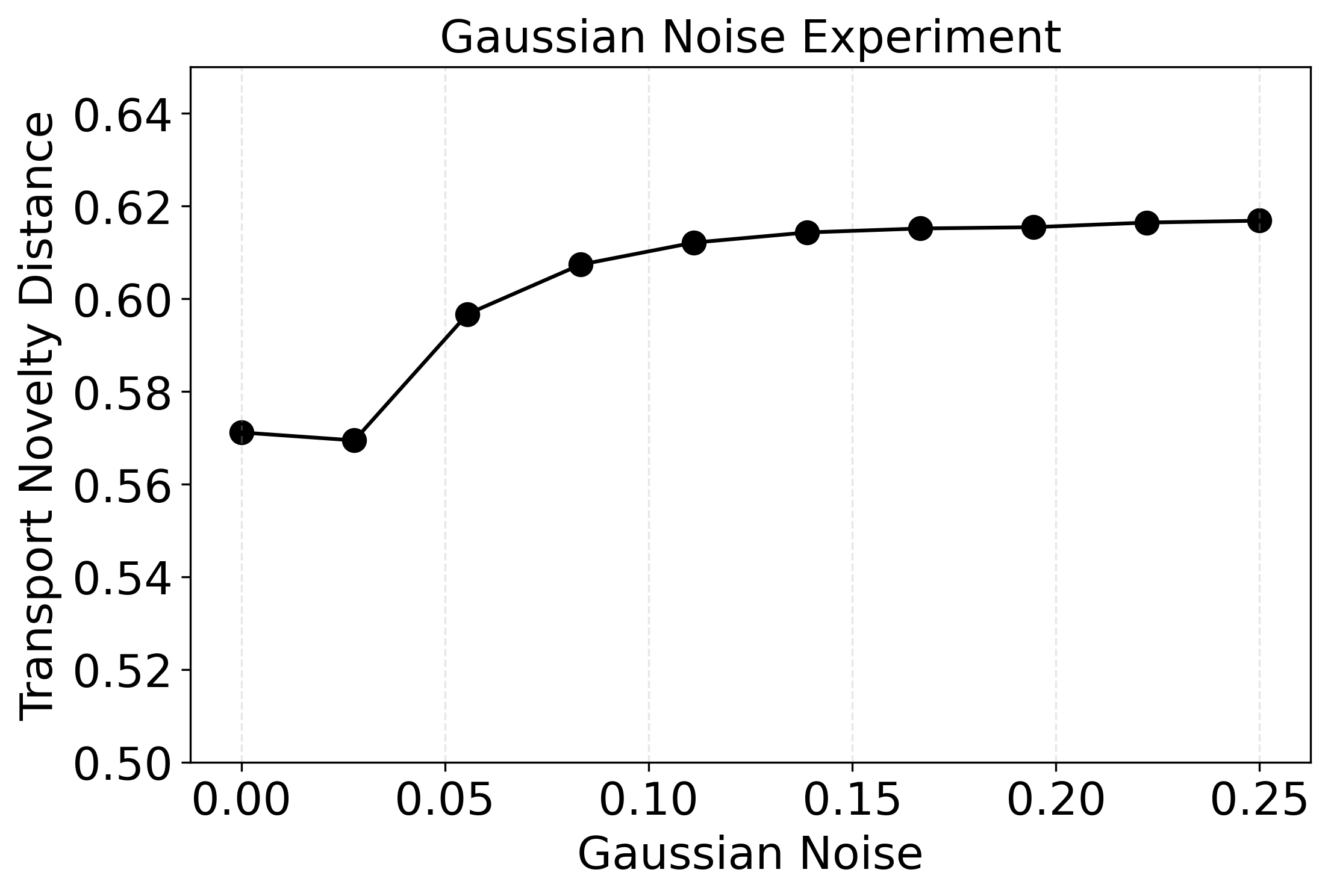}
        \caption{Gaussian noise added to $(x,y,z)$ of materials and the corresponding TNovD.}
        \label{fig:gaussian_noise}

    \end{subfigure}
    
    \vskip\baselineskip  

    \begin{subfigure}[b]{0.45\textwidth}
        \centering
        \includegraphics[width=\textwidth]{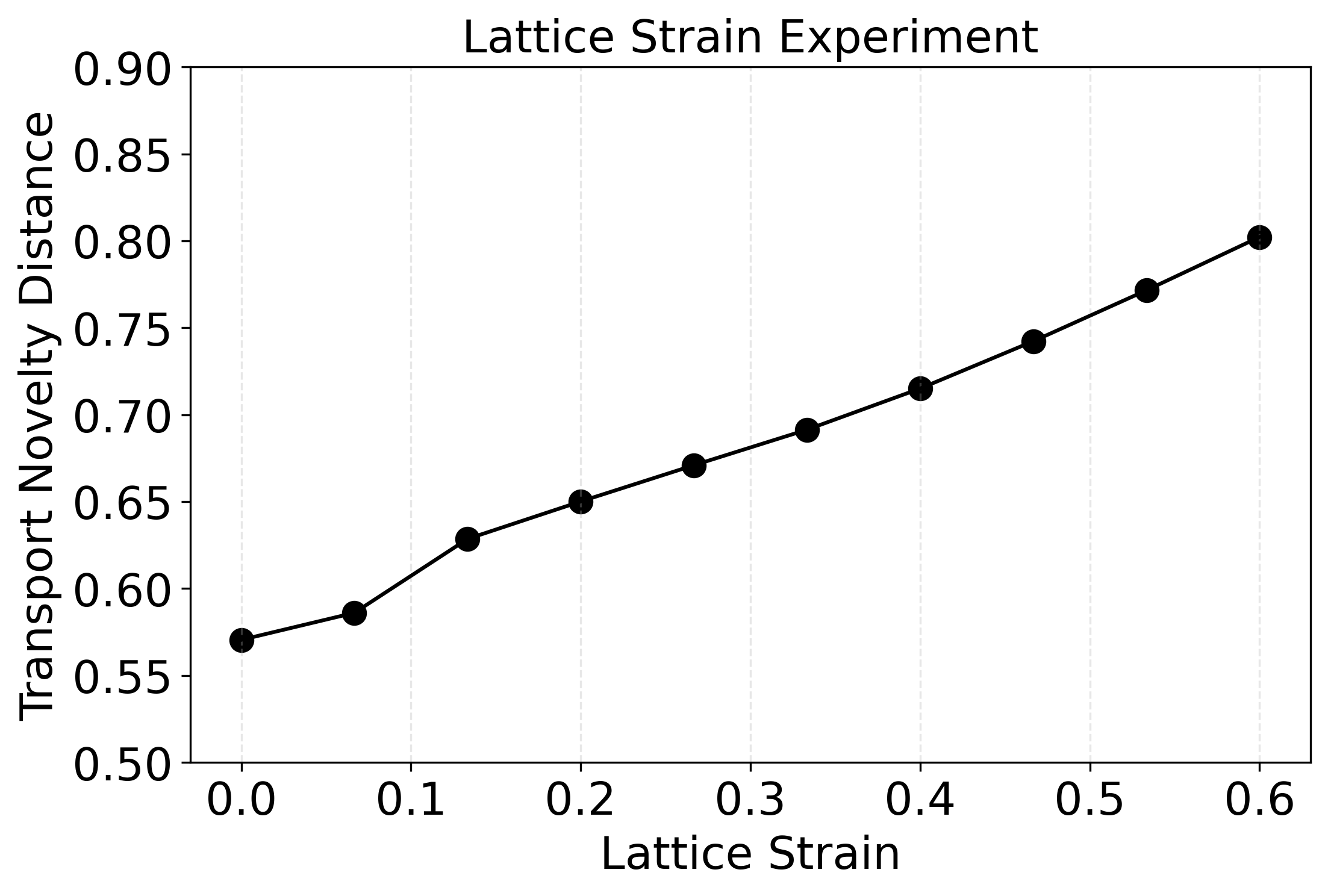}
        \caption{Random Lattice shear of materials and the corresponding TNovD.}
        \label{fig:latti}
    \end{subfigure}
    \hfill
    \begin{subfigure}[b]{0.45\textwidth}
        \centering
        \includegraphics[width=\textwidth]{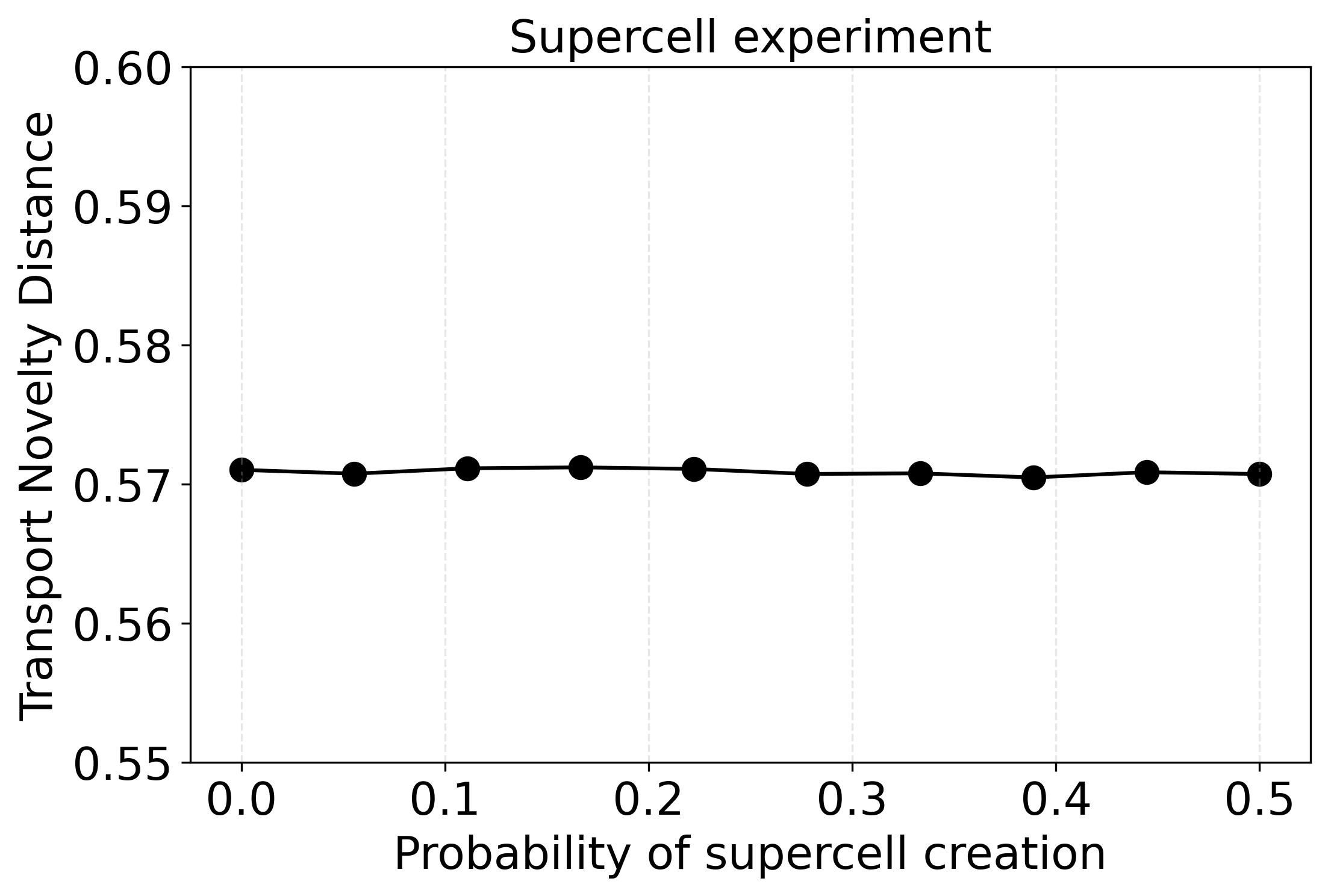}
        \caption{Random supercells of materials and the corresponding TNovD.}
        \label{fig:supercell}
    \end{subfigure}

    \begin{subfigure}[b]{0.45\textwidth}
        \centering
        \includegraphics[width=\textwidth]{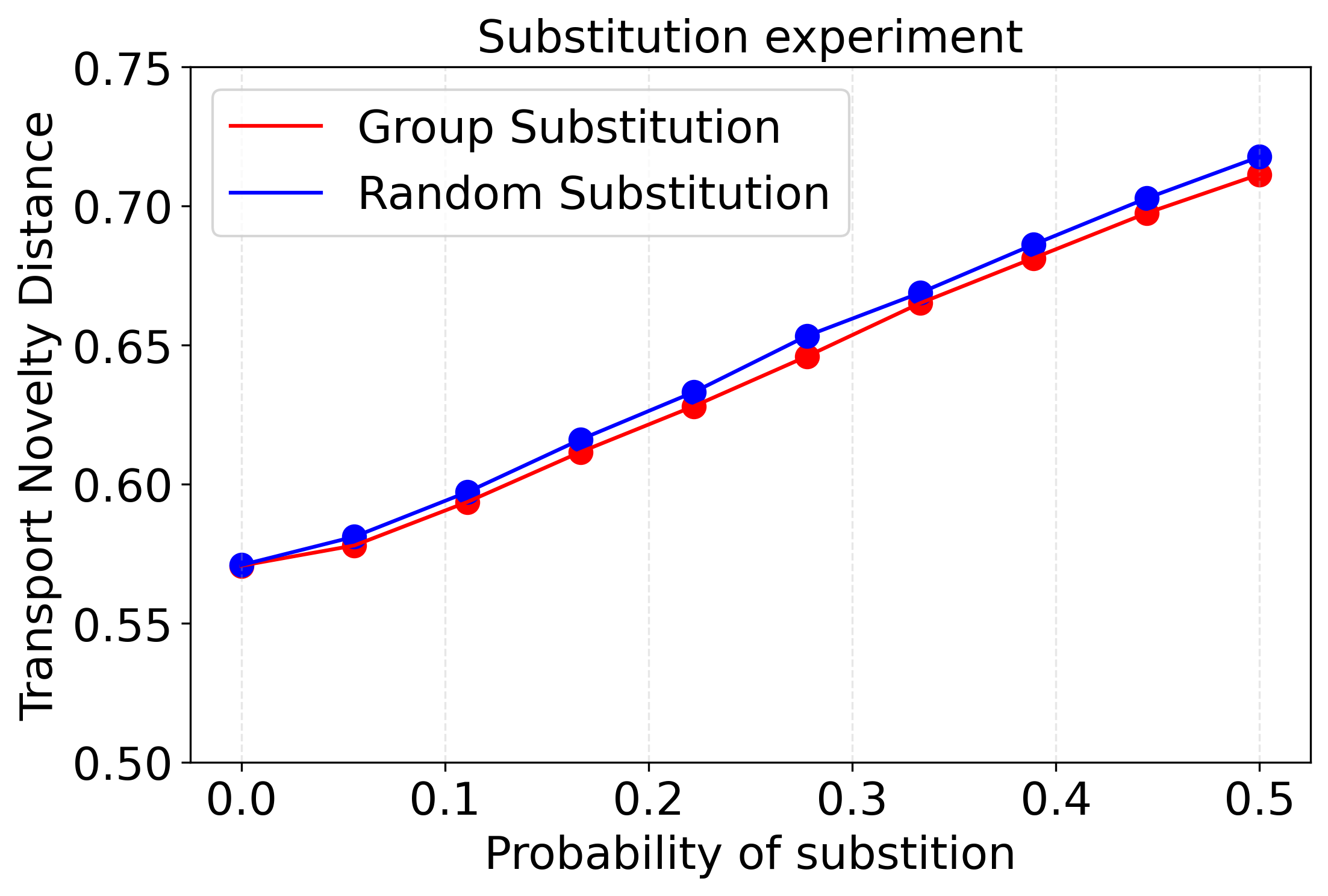}
        \caption{In-group/Random Substitutions of atoms and the corresponding TNovD.}
        \label{fig:substi}
    \end{subfigure}

    \caption{Different toy experiments performed to test the TNovD behavior for specific test cases.}
\label{fig:novelty_summary}
\end{figure}

The result for every iteration of every corruption is shown in Fig. \ref{fig:novelty_summary}. 

In the Fig. \ref{fig:novelty_summary} (a), we observe that our TNovD performs excellently in recognizing memorized data, with an almost perfectly linear increase for increasing amounts of leaked training data. 
For the Gaussian noise on $(x,y,z)$ (Fig. \ref{fig:novelty_summary} (b)), our TNovD stays nearly identical for the first addition of noise, before it steadily increases for several iterations until reaching a plateau. The small drop of the TNovD after the first iteration at a Gaussian noise of 0.2 is explained by the contradicting behavior of memorization and quality regime that form the final TNovD. While the values of the memorization regime decrease sharply, the observed increase in the quality regime is comparably smaller. However, with a higher standard deviation $\sigma$, we see that our TNovD quickly picks up on the low-quality structure of the resulting materials, and the high values of the quality regime are dominating the final TNovD. Fig. \ref{fig:novelty_summary} (c) displays gradually increasing TNovD values with increasing amounts of lattice strain. This shows that our TNovD identifies lattice deformation even for small values of lattice strain. Further, we also see that our model is perfectly invariant to supercells (Fig. \ref{fig:novelty_summary} (d)). Fig. \ref{fig:novelty_summary} (e) displays the TNovD values for increasing probabilities of random substitutions, as well as in-group substitutions. We observe that the TNovD correctly identifies the results of both substitutions as lower quality materials. It thereby penalizes random substitutions slightly more strongly than in-group ones. 

\paragraph{Behavior on the WBM-Dataset}
In addition to tests performed on the validation dataset, we used the WBM data published by \cite{wbmdataset}. They created novel crystal structures by substituting one atom in the structure with another, chemically similar atom. Overall, they performed five steps of substitution, increasing the number of swapped atoms by one in every iteration. Stability of the resulting crystal structures was determined by DFT calculations. As a result, five individual datasets were published, each iteration increasingly different (but probably also less physically realistic) than the original data taken from the Materials Project (\cite{materialsproject}). As the training data set (the one we benchmark the steps against) we choose the training data set of MP20. 
For every dataset, all stable structures with energy above hull values below 0 were extracted, and outliers (structures with energy above hull values of -10) as well as double entries were removed. Then, the TNovD was calculated for every dataset individually, as well as for the complete MP20 data (train, validation, and test set), which we used as a proxy for step $0$. Note that we treated the MP20 data as the basis for substitution, although \cite{wbmdataset} used the whole Materials Project data as input.  

\begin{figure}[htpb]
    \centering
    \includegraphics[width=0.8\textwidth]{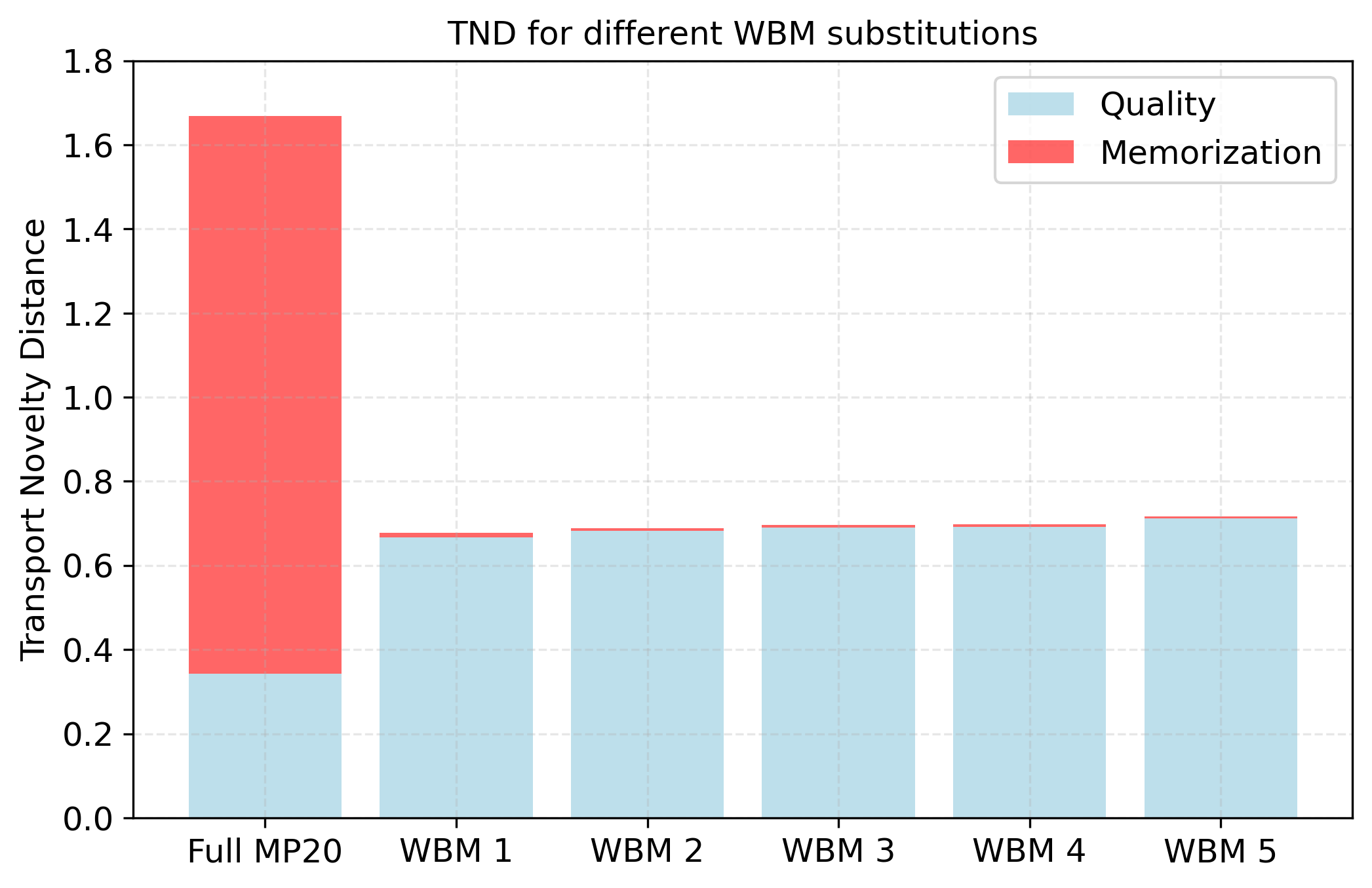}
    \caption{Transport Novelty Distance for the full MP20 data (train + validation + test set) and all five steps of substitution of the WBM dataset. Quality (blue) and memorization (red) components of the final TNovD are shown separately for each model.}
    \label{fig:wbm_scores}
\end{figure}

The total TNovD values are presented in Fig. \ref{fig:wbm_scores}, showing the individual memorization and validity components. The high values of the full MP20 dataset are mostly a result of the large memorization component. Since the complete MP20 dataset includes the MP20 train split, which was used as a comparison for the generated (here: substituted) data, identical structures are present in both datasets. Fig. \ref{fig:wbm_scores} shows that the memorization component detects this. 

From WBM 1 onward, the memorization component continuously decreases, though its values are generally very small. This is due to the fact that the crystal structures get more and more dissimilar to the MP20 train split, since more atoms are replaced at every step. Likewise, values for the quality component increase slightly with increasing atom substitutions, since the resulting structures become more dissimilar from the MP20 train split. Overall, the WBM 1 dataset achieves the best TNovD score, as no memorization is detected and quality is good. The total TNovD value of the WBM 1 data set (0.68) is larger than the baseline of 0.57 in the toy experiments (Fig. \ref{fig:novelty_summary}). Since the WBM structures were created by substituting the full data of the Materials Project, the initial structures already deviate much more from the MP20 train split than those of the MP20 validation split used for the toy examples above. Therefore, the feature distributions of all WBM steps are further away from the MP20 train set and the final TNovD value is accordingly larger.  

The negligible contribution of the memorization component to the TNovD suggests that as soon as one atom is replaced, our featurizer already sees this as a new material, which is in accordance with how we trained it. Additionally, to be in line with most generative models, we treated the distribution of the MP20 train split as the input for substitution, although in reality, the complete Materials Project data was used. Assumably, the original structures of many substituted materials are not part of the MP20 train set and memorization is not detected correctly.

\paragraph{Transport Novelty Distance for common generative models}
In addition to the different tests shown above, we calculated the TNovD for common generative models and benchmarked them.  We used the same crystal structures as \cite{negishi2025continuousuniquenessnoveltymetrics}, who generated an identical amount of 10,000 structures by training the six models \cite{joshi2025allatomdiffusiontransformersunified, MatterGen2025, xie2022crystal, jiao2024crystalstructurepredictionjoint, jiao2024spacegroupconstrainedcrystal, llmcrystalgeneration} on the MP20 train split. 

\begin{figure}[htpb]
    \centering
    \includegraphics[width=0.8\textwidth]{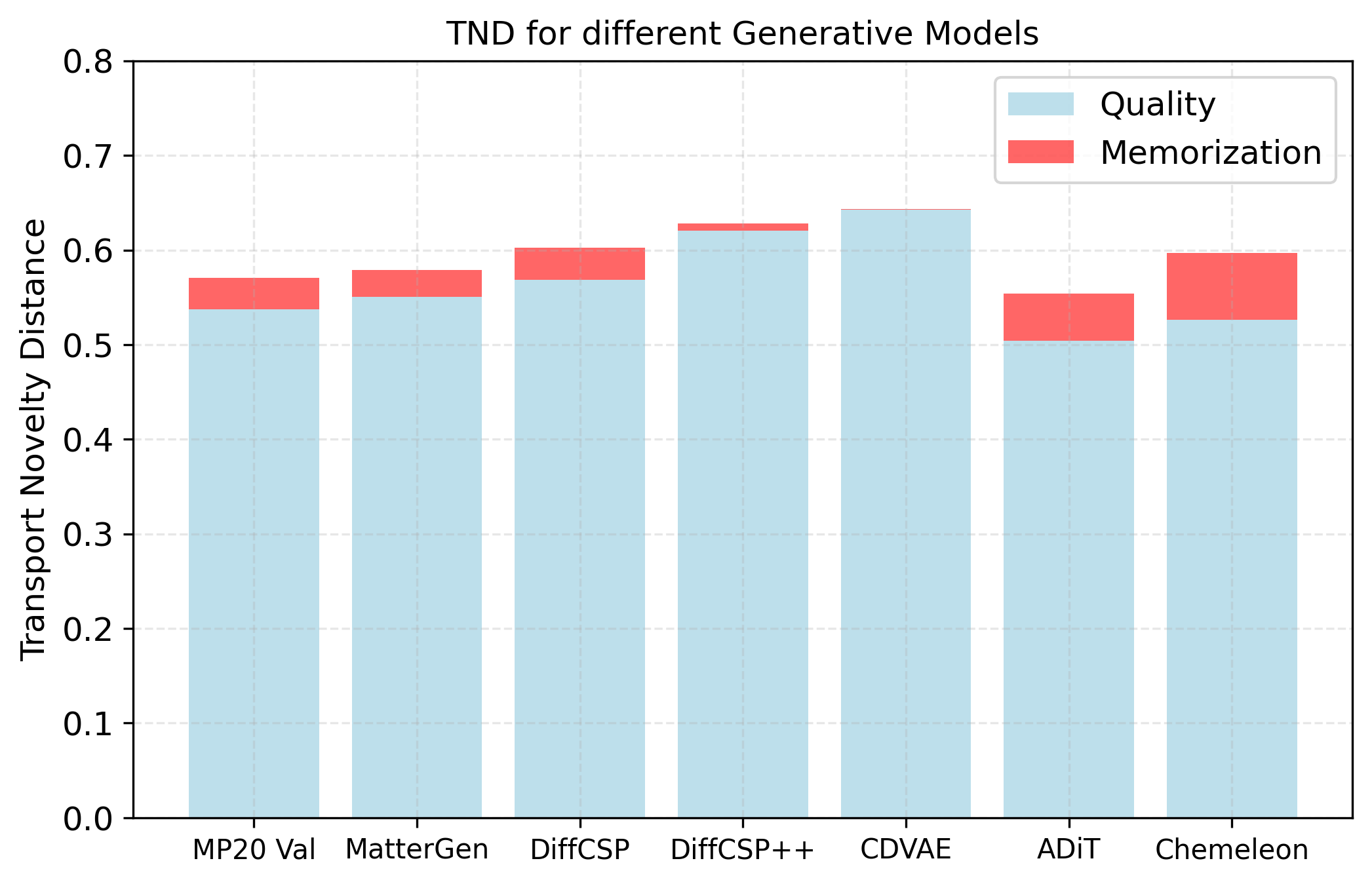}
    \caption{TNovD for the MP20 validation set, as well as for 10,000 predicted samples from six different generative models, all trained on the same MP20 train split. Quality (blue) and memorization (red) components of the final TNovD are shown separately for each model.}
    \label{fig:model_eval}
\end{figure}

The generated crystal structures were embedded with our pretrained invariant GNN. These features were then used to calculate the TNovD with respect to the MP20 train set. 
The bars in Fig. \ref{fig:model_eval} display the TNovD for each generative model. We split up the \emph{total TNovD} into the \emph{quality} (blue) and the \emph{memorization} (red) component. Additionally, we added the values for the MP20 validation set for comparison. The best performing "models" are ADiT \cite{joshi2025allatomdiffusiontransformersunified}, then the validation set, closely followed by MatterGen \cite{MatterGen2025}. 

Furthermore, DiffCSP outperforms DiffCSP++, which aligns with the MSUN (meta-stable, unique, novel) results in \cite{duval2025lematgenbench}, but contradicts the distributional ones. The oldest model, CDVAE \cite{xie2022crystal}, is also the worst-performing one. Its generated structures do not match the training structures, resulting in no memorization but a bad quality score. Further, Chemeleon \cite{llmcrystalgeneration} seems to memorize the most.  

Interestingly, ADiT leads due to achieving the best results in the quality component. We hypothesize that ADiT memorizes more, but in a way that is not too heavily penalized by our TNovD. This could be due to it memorizing \emph{but not exactly reproducing}, which is not perfectly picked up by our metric. This seems in line with the printed novelty and coverage scores in the feature space, which indicate that ADiT has a very good coverage score but a low novelty score, explaining both the quality and memorization values of the TNovD. MatterGen and the validation set have similar coverage and memorization scores, underlining similar performances. We remark that it is not clear that our featurizer accurately captures stability properties; therefore, these rankings should be taken with a grain of salt. We leave exploration for stability-aware distributional evaluation of models for future work. This could include the consideration of energies above hull and/or dynamic stability as extracted from harmonic phonon calculations. Further considerations, such as synthesizability measures, might also be included in future work (e.g. \cite{metni2025generativemodelscrystallinematerials, park2025closingsynt, D4DD00394B}.

\section*{Conclusion}

Using generative models to create potentially novel materials is an emerging field in materials science. Although numerous models with varying architectural designs have been developed, methods to validate these models by combining novelty and quality within a single framework are relatively sparse and often limited to an instance level of individual structures. In this paper, we developed an evaluation metric, the \emph{Transport Novelty Distance}. It is based on the Wasserstein distance, which compares the abstract crystal structure features of the generative models training set with those of the newly generated material. It lays out a modular blueprint for assessing quality and novelty in a combined framework. However, several improvements, such as those related to hyperparameter tuning and stability, are possible. 

In our results, we see that the novelty component of the TNovD contributes significantly less than the quality component. This is not surprising, as for generative models, it is a very \emph{unlikely} event to map precisely to a training data point. We would like to emphasize here that our novelty description depends on our training dataset (MP20), the chosen augmentations (e.g., supercells), and the embedding of crystalline materials. 
Based on our current materials representation, the TNovD results suggest that fitting models to the correct data distribution is the major challenge of material generative models. The low memorization component indicates the prediction of novel structures, and a quality component close to the validation data suggests high quality materials. However, our metric does not capture the materials stability directly. 
The incorporation of a pre-trained energy model, such as MACE \cite{Batatia2023AFM}, could be fruitful for future work, as well as an extension for conditional generative models. Furthermore, for specific applications, the set of positive and negative examples in the InfoNCE objective can be improved, for example, by explicitly considering disorder within the structural embedding.

Overall, we note that although our metric is introduced for specific material generative models, extensions to other domains, such as image or molecule generative models, should be possible. We see this as an increasingly important direction of development, as generative models advance and hence tend to memorize more.  

\section*{Acknowledgements}
J. G. was supported by ERC Grant MultiBonds (grant agreement no. 101161771; funded by the European Union. Views and opinions expressed are, however, those of the author(s) only and do not necessarily reflect those of the European Union or the European Research Council Executive Agency. Neither the European Union nor the granting authority can be held responsible for them.).

\bibliography{bibi}

@article{peyre2019,
year = {2019},
volume = {11},
journal = {Foundations and Trends® in Machine Learning},
title = {Computational Optimal Transport: With Applications to Data Science},
doi = {10.1561/2200000073},
issn = {1935-8237},
number = {5-6},
pages = {355-607},
author = {Gabriel Peyré and Marco Cuturi}
}

@InProceedings{satorras,
  title = 	 {E(n) Equivariant Graph Neural Networks},
  author =       {Satorras, V\'{\i}ctor Garcia and Hoogeboom, Emiel and Welling, Max},
  booktitle = 	 {Proceedings of the 38th International Conference on Machine Learning},
  pages = 	 {9323--9332},
  year = 	 {2021},
  editor = 	 {Meila, Marina and Zhang, Tong},
  volume = 	 {139},
  series = 	 {Proceedings of Machine Learning Research},
  month = 	 {18--24 Jul},
  publisher =    {PMLR},
  url = 	 {https://proceedings.mlr.press/v139/satorras21a.html}
}

@article{batatia2023AFM,
  title={A foundation model for atomistic materials chemistry},
  author={Batatia, Ilyes and Benner, Philipp and Chiang, Yuan and Elena, Alin M and Kov{\'a}cs, D{\'a}vid P and Riebesell, Janosh and Advincula, Xavier R and Asta, Mark and Avaylon, Matthew and Baldwin, William J and others},
  journal={The Journal of chemical physics},
  volume={163},
  number={18},
  year={2025},
  publisher={AIP Publishing}
}

@article{tameling2021colocalization,
  title={Colocalization for super-resolution microscopy via optimal transport},
  author={Tameling, Carla and Stoldt, Stefan and Stephan, Till and Naas, Julia and Jakobs, Stefan and Munk, Axel},
  journal={Nature computational science},
  volume={1},
  number={3},
  pages={199--211},
  year={2021},
  publisher={Nature Publishing Group US New York}
}

@article{pot_package,
  author  = {R{\'e}mi Flamary and Nicolas Courty and Alexandre Gramfort and Mokhtar Z. Alaya and AurÃ©lie Boisbunon and Stanislas Chambon and Laetitia Chapel and Adrien Corenflos and Kilian Fatras and Nemo Fournier and LÃ©o Gautheron and Nathalie T.H. Gayraud and Hicham Janati and Alain Rakotomamonjy and Ievgen Redko and Antoine Rolet and Antony Schutz and Vivien Seguy and Danica J. Sutherland and Romain Tavenard and Alexander Tong and Titouan Vayer},
  title   = {POT: Python Optimal Transport},
  journal = {Journal of Machine Learning Research},
  year    = {2021},
  volume  = {22},
  number  = {78},
  pages   = {1--8},
  url     = {http://jmlr.org/papers/v22/20-451.html}
}

@article{Klipfel_Fregier_Sayede_Bouraoui_2024, title={Vector Field Oriented Diffusion Model for Crystal Material Generation}, volume={38}, url={https://ojs.aaai.org/index.php/AAAI/article/view/30224}, DOI={10.1609/aaai.v38i20.30224}, number={20}, journal={Proceedings of the AAAI Conference on Artificial Intelligence}, author={Klipfel, Astrid and Fregier, Yaël and Sayede, Adlane and Bouraoui, Zied}, year={2024}, month={Mar.}, pages={22193-22201} }

@inproceedings{pytorchpaper,
 author = {Paszke, Adam and Gross, Sam and Massa, Francisco and Lerer, Adam and Bradbury, James and Chanan, Gregory and Killeen, Trevor and Lin, Zeming and Gimelshein, Natalia and Antiga, Luca and Desmaison, Alban and Kopf, Andreas and Yang, Edward and DeVito, Zachary and Raison, Martin and Tejani, Alykhan and Chilamkurthy, Sasank and Steiner, Benoit and Fang, Lu and Bai, Junjie and Chintala, Soumith},
 booktitle = {Advances in Neural Information Processing Systems},
 editor = {H. Wallach and H. Larochelle and A. Beygelzimer and F. d\textquotesingle Alch\'{e}-Buc and E. Fox and R. Garnett},
 pages = {},
 publisher = {Curran Associates, Inc.},
 title = {PyTorch: An Imperative Style, High-Performance Deep Learning Library},
 url = {https://proceedings.neurips.cc/paper_files/paper/2019/file/bdbca288fee7f92f2bfa9f7012727740-Paper.pdf},
 volume = {32},
 year = {2019}
}

@inproceedings{Fey2019,
  title={Fast Graph Representation Learning with {PyTorch Geometric}},
  author={Fey, Matthias and Lenssen, Jan E.},
  booktitle={ICLR Workshop on Representation Learning on Graphs and Manifolds},
  year={2019},
}

@inproceedings{Fey2025,
  title={{PyG} 2.0: Scalable Learning on Real World Graphs},
  author={Fey, Matthias and Sunil, Jinu and Nitta, Akihiro and Puri, Rishi and Shah, Manan and Stojanovi{\v{c}}, Bla{\v{z}} and Bendias, Ramona and Barghi, Alexandria and Kocijan, Vid and Zhang, Zecheng and He, Xinwei and Lenssen, Jan E. and Leskovec, Jure},
  booktitle={Temporal Graph Learning Workshop @ KDD},
  year={2025}
}

@article{Ong2012b,
author = {Ong, Shyue Ping and Richards, William Davidson and Jain, Anubhav and Hautier, Geoffroy and Kocher, Michael and Cholia, Shreyas and Gunter, Dan and Chevrier, Vincent L. and Persson, Kristin A. and Ceder, Gerbrand},

issn = {09270256},
journal = {Computational Materials Science},
month = feb,
pages = {314--319},
title = {{Python Materials Genomics (pymatgen): A robust, open-source python library for materials analysis}},
url = {http://linkinghub.elsevier.com/retrieve/pii/S0927025612006295},
volume = {68},
year = {2013}
}

@inproceedings{
yoon2023diffusion,
title={Diffusion Probabilistic Models Generalize when They Fail to Memorize},
author={TaeHo Yoon and Joo Young Choi and Sehyun Kwon and Ernest K. Ryu},
booktitle={ICML 2023 Workshop on Structured Probabilistic Inference {\&} Generative Modeling},
year={2023},
url={https://openreview.net/forum?id=shciCbSk9h}
}

@article{buchanan2025edgememorizationdiffusionmodels,
      title={On the Edge of Memorization in Diffusion Models}, 
      author={Sam Buchanan and Druv Pai and Yi Ma and Valentin De Bortoli},
      year={2025},
      journal={arXiv preprint arXiv:2508.17689}
}

@inbook{MarinoGerolinNenna2017,
  title        = {Optimal transportation theory with repulsive costs},
  booktitle    = {Topological Optimization and Optimal Transport},
  author       = {Simone Di Marino and Augusto Gerolin and Luca Nenna},
  editor       = {Ma{\"i}tine Bergounioux and {\'E}douard Oudet and Martin Rumpf and Guillaume Carlier and Thierry Champion and Filippo Santambrogio},
  publisher    = {De Gruyter},
  address      = {Berlin, Boston},
  pages        = {204--256},
  year         = {2017},
  doi          = {10.1515/9783110430417-010},
  url          = {https://doi.org/10.1515/9783110430417-010},
  isbn         = {9783110430417}
}

@inproceedings{
dinh2017density,
title={Density estimation using Real {NVP}},
author={Laurent Dinh and Jascha Sohl-Dickstein and Samy Bengio},
booktitle={International Conference on Learning Representations},
year={2017},
url={https://openreview.net/forum?id=HkpbnH9lx}
}

@inproceedings{
song2021scorebased,
title={Score-Based Generative Modeling through Stochastic Differential Equations},
author={Yang Song and Jascha Sohl-Dickstein and Diederik P Kingma and Abhishek Kumar and Stefano Ermon and Ben Poole},
booktitle={International Conference on Learning Representations},
year={2021},
url={https://openreview.net/forum?id=PxTIG12RRHS}
}

@inproceedings{onken2021otflow, 
    title={{OT-Flow}: Fast and Accurate Continuous Normalizing Flows via Optimal Transport},
    author={Derek Onken and Samy Wu Fung and Xingjian Li and Lars Ruthotto},
    volume={35}, 
    number={10}, 
    booktitle={AAAI Conference on Artificial Intelligence}, 
    year={2021}, 
    month={May},
    pages={9223--9232},
    url={https://ojs.aaai.org/index.php/AAAI/article/view/17113}, 
}

@inbook{villani,
author = {Villani, Cédric},
year = {2008},
month = {01},
pages = {xxii+973},
title = {Optimal transport -- Old and new},
volume = {338},
doi = {10.1007/978-3-540-71050-9}
}

@article{
tong2024improving,
title={Improving and generalizing flow-based generative models with minibatch optimal transport},
author={Alexander Tong and Kilian Fatras and Nikolay Malkin and Guillaume Huguet and Yanlei Zhang and Jarrid Rector-Brooks and Guy Wolf and Yoshua Bengio},
journal={Transactions on Machine Learning Research},
issn={2835-8856},
year={2024},
url={https://openreview.net/forum?id=CD9Snc73AW},
note={Expert Certification}
}

@article{ye2025provableseparationsmemorizationgeneralization,
      title={Provable Separations between Memorization and Generalization in Diffusion Models}, 
      author={Zeqi Ye and Qijie Zhu and Molei Tao and Minshuo Chen},
      year={2025},
      journal={arXiv preprint arXiv:2511.03202}
}

@inproceedings{
pidstrigach2022scorebased,
title={Score-Based Generative Models Detect Manifolds},
author={Jakiw Pidstrigach},
booktitle={Advances in Neural Information Processing Systems},
editor={Alice H. Oh and Alekh Agarwal and Danielle Belgrave and Kyunghyun Cho},
year={2022},
url={https://openreview.net/forum?id=AiNrnIrDfD9}
}

@InProceedings{WyckoffDiff,
  title = 	 {{W}yckoff{D}iff – A Generative Diffusion Model for Crystal Symmetry},
  author =       {Ekstr\"{o}m Kelvinius, Filip and Andersson, Oskar B. and Parackal, Abhijith S and Qian, Dong and Armiento, Rickard and Lindsten, Fredrik},
  booktitle = 	 {Proceedings of the 42nd International Conference on Machine Learning},
  pages = 	 {15130--15147},
  year = 	 {2025},
  editor = 	 {Singh, Aarti and Fazel, Maryam and Hsu, Daniel and Lacoste-Julien, Simon and Berkenkamp, Felix and Maharaj, Tegan and Wagstaff, Kiri and Zhu, Jerry},
  volume = 	 {267},
  series = 	 {Proceedings of Machine Learning Research},
  month = 	 {13--19 Jul},
  publisher =    {PMLR},
  url = 	 {https://proceedings.mlr.press/v267/ekstrom-kelvinius25a.html},
}

@article{baird2024,
  title={matbench-genmetrics: A Python library for benchmarking crystal structure generative models using time-based splits of Materials Project structures},
  author={Baird, Sterling G and Sayeed, Hasan M and Montoya, Joseph and Sparks, Taylor D},
  journal={Journal of Open Source Software},
  volume={9},
  number={97},
  pages={5618},
  year={2024}
}

@article{zhang2024,
  title={An interpretable evaluation of entropy-based novelty of generative models},
  author={Zhang, Jingwei and Li, Cheuk Ting and Farnia, Farzan},
  journal={arXiv preprint arXiv:2402.17287},
  year={2024}
}

@article{sanchez-lengeling2021a,
  author = {Sanchez-Lengeling, Benjamin and Reif, Emily and Pearce, Adam and Wiltschko, Alexander B.},
  title = {A Gentle Introduction to Graph Neural Networks},
  journal = {Distill},
  year = {2021},
  note = {https://distill.pub/2021/gnn-intro},
  doi = {10.23915/distill.00033}
}

@article{PhysRevLett.120.145301,
  title = {Crystal Graph Convolutional Neural Networks for an Accurate and Interpretable Prediction of Material Properties},
  author = {Xie, Tian and Grossman, Jeffrey C.},
  journal = {Phys. Rev. Lett.},
  volume = {120},
  issue = {14},
  pages = {145301},
  numpages = {6},
  year = {2018},
  month = {Apr},
  publisher = {American Physical Society},
  doi = {10.1103/PhysRevLett.120.145301},
  url = {https://link.aps.org/doi/10.1103/PhysRevLett.120.145301}
}

@inproceedings{schnet,
 author = {Sch\"{u}tt, Kristof and Kindermans, Pieter-Jan and Sauceda Felix, Huziel Enoc and Chmiela, Stefan and Tkatchenko, Alexandre and M\"{u}ller, Klaus-Robert},
 booktitle = {Advances in Neural Information Processing Systems},
 editor = {I. Guyon and U. Von Luxburg and S. Bengio and H. Wallach and R. Fergus and S. Vishwanathan and R. Garnett},
 pages = {},
 publisher = {Curran Associates, Inc.},
 title = {SchNet: A continuous-filter convolutional neural network for modeling quantum interactions},
 url = {https://proceedings.neurips.cc/paper_files/paper/2017/file/303ed4c69846ab36c2904d3ba8573050-Paper.pdf},
 volume = {30},
 year = {2017}
}

@InProceedings{pmlr-v70-gilmer17a,
  title = 	 {Neural Message Passing for Quantum Chemistry},
  author =       {Justin Gilmer and Samuel S. Schoenholz and Patrick F. Riley and Oriol Vinyals and George E. Dahl},
  booktitle = 	 {Proceedings of the 34th International Conference on Machine Learning},
  pages = 	 {1263--1272},
  year = 	 {2017},
  editor = 	 {Precup, Doina and Teh, Yee Whye},
  volume = 	 {70},
  series = 	 {Proceedings of Machine Learning Research},
  month = 	 {06--11 Aug},
  publisher =    {PMLR},
  url = 	 {https://proceedings.mlr.press/v70/gilmer17a.html}
}

@ARTICLE{Scarselli,
  author={Scarselli, Franco and Gori, Marco and Tsoi, Ah Chung and Hagenbuchner, Markus and Monfardini, Gabriele},
  journal={IEEE Transactions on Neural Networks}, 
  title={The Graph Neural Network Model}, 
  year={2009},
  volume={20},
  number={1},
  pages={61-80},
  doi={10.1109/TNN.2008.2005605}}

@article{oord2019representationlearningcontrastivepredictive,
      title={Representation Learning with Contrastive Predictive Coding}, 
      author={Aaron van den Oord and Yazhe Li and Oriol Vinyals},
      year={2018},
      journal={arXiv preprint arXiv:1807.03748} 
}

@article{szymanski2025establishingbaselinesgenerativediscovery,
      title={Establishing baselines for generative discovery of inorganic crystals}, 
      author={Nathan J. Szymanski and Christopher J. Bartel},
      year={2025},
      journal={arXiv preprint arXiv:2501.02144}
}

@inproceedings{
jeon2025understanding,
title={Understanding and Mitigating Memorization in Generative Models via Sharpness of Probability Landscapes},
author={Dongjae Jeon and Dueun Kim and Albert No},
booktitle={Forty-second International Conference on Machine Learning},
year={2025},
url={https://openreview.net/forum?id=EW2JR5aVLm}
}

@inproceedings{
jiralerspong2023feature,
title={Feature Likelihood Score: Evaluating the Generalization of Generative Models Using Samples},
author={Marco Jiralerspong and Joey Bose and Ian Gemp and Chongli Qin and Yoram Bachrach and Gauthier Gidel},
booktitle={Thirty-seventh Conference on Neural Information Processing Systems},
year={2023},
url={https://openreview.net/forum?id=l2VKZkolT7}
}

@inproceedings{stein2023,
 author = {Stein, George and Cresswell, Jesse and Hosseinzadeh, Rasa and Sui, Yi and Ross, Brendan and Villecroze, Valentin and Liu, Zhaoyan and Caterini, Anthony L and Taylor, Eric and Loaiza-Ganem, Gabriel},
 booktitle = {Advances in Neural Information Processing Systems},
 editor = {A. Oh and T. Naumann and A. Globerson and K. Saenko and M. Hardt and S. Levine},
 pages = {3732--3784},
 publisher = {Curran Associates, Inc.},
 title = {Exposing flaws of generative model evaluation metrics and their unfair treatment of diffusion models},
 url = {https://proceedings.neurips.cc/paper_files/paper/2023/file/0bc795afae289ed465a65a3b4b1f4eb7-Paper-Conference.pdf},
 volume = {36},
 year = {2023}
}

@article{Usman_akbar2025,
doi = {10.1088/2632-2153/ad9a3a},
url = {https://doi.org/10.1088/2632-2153/ad9a3a},
year = {2025},
month = {jan},
publisher = {IOP Publishing},
volume = {6},
number = {1},
pages = {015022},
author = {Usman Akbar, Muhammad and Wang, Wuhao and Eklund, Anders},
title = {Beware of diffusion models for synthesizing medical images—a comparison with GANs in terms of memorizing brain MRI and chest x-ray images},
journal = {Machine Learning: Science and Technology}
}

@Article{onwuli2023,
author ="Onwuli, Anthony and Hegde, Ashish V. and Nguyen, Kevin V. T. and Butler, Keith T. and Walsh, Aron",
title  ="Element similarity in high-dimensional materials representations",
journal  ="Digital Discovery",
year  ="2023",
volume  ="2",
issue  ="5",
pages  ="1558-1564",
publisher  ="RSC",
doi  ="10.1039/D3DD00121K",
url  ="http://dx.doi.org/10.1039/D3DD00121K"}

@article{ChemEMD,
author = {Hargreaves, Cameron J. and Dyer, Matthew S. and Gaultois, Michael W. and Kurlin, Vitaliy A. and Rosseinsky, Matthew J.},
title = {The Earth Mover’s Distance as a Metric for the Space of Inorganic Compositions},
journal = {Chemistry of Materials},
volume = {32},
number = {24},
pages = {10610-10620},
year = {2020},
doi = {10.1021/acs.chemmater.0c03381},

URL = {
        https://doi.org/10.1021/acs.chemmater.0c03381
},eprint = { 
    
        https://doi.org/10.1021/acs.chemmater.0c03381}

}

@inproceedings{fid,
 author = {Heusel, Martin and Ramsauer, Hubert and Unterthiner, Thomas and Nessler, Bernhard and Hochreiter, Sepp},
 booktitle = {Advances in Neural Information Processing Systems},
 editor = {I. Guyon and U. Von Luxburg and S. Bengio and H. Wallach and R. Fergus and S. Vishwanathan and R. Garnett},
 pages = {},
 publisher = {Curran Associates, Inc.},
 title = {GANs Trained by a Two Time-Scale Update Rule Converge to a Local Nash Equilibrium},
 url = {https://proceedings.neurips.cc/paper_files/paper/2017/file/8a1d694707eb0fefe65871369074926d-Paper.pdf},
 volume = {30},
 year = {2017}
}

@article{farghly2025diffusionmodelsmanifoldhypothesis,
      title={Diffusion Models and the Manifold Hypothesis: Log-Domain Smoothing is Geometry Adaptive}, 
      author={Tyler Farghly and Peter Potaptchik and Samuel Howard and George Deligiannidis and Jakiw Pidstrigach},
      year={2025},
      journal={arXiv preprint arXiv:2510.02305}
}

@article{
scarvelis2025closedform,
title={Closed-Form Diffusion Models},
author={Christopher Scarvelis and Haitz S{\'a}ez de Oc{\'a}riz Borde and Justin Solomon},
journal={Transactions on Machine Learning Research},
issn={2835-8856},
year={2025},
url={https://openreview.net/forum?id=JkMifr17wc},
note={}
}

@article{koker2022graphcontrastivelearningmaterials,
      title={Graph Contrastive Learning for Materials}, 
      author={Teddy Koker and Keegan Quigley and Will Spaeth and Nathan C. Frey and Lin Li},
      year={2022},
      journal={arXiv preprint arXiv:2211.13408}, 
}

@inproceedings{tack2020csi,
  title={CSI: Novelty Detection via Contrastive Learning on Distributionally Shifted Instances},
  author={Jihoon Tack and Sangwoo Mo and Jongheon Jeong and Jinwoo Shin},
  booktitle={Advances in Neural Information Processing Systems},
  year={2020}
}

@article{1991TheCI,
author = {Hall, S. R. and Allen, F. H. and Brown, I. D.},
title = {The crystallographic information file (CIF): a new standard archive file for crystallography},
journal = {Acta Crystallographica Section A},
volume = {47},
number = {6},
pages = {655-685},
doi = {https://doi.org/10.1107/S010876739101067X},
url = {https://onlinelibrary.wiley.com/doi/abs/10.1107/S010876739101067X},
eprint = {https://onlinelibrary.wiley.com/doi/pdf/10.1107/S010876739101067X},
year = {1991}
}

@inproceedings{
kipf2017semisupervised,
title={Semi-Supervised Classification with Graph Convolutional Networks},
author={Thomas N. Kipf and Max Welling},
booktitle={International Conference on Learning Representations},
year={2017},
url={https://openreview.net/forum?id=SJU4ayYgl}
}

@article{negishi2025continuousuniquenessnoveltymetrics,
      title={Continuous Uniqueness and Novelty Metrics for Generative Modeling of Inorganic Crystals}, 
      author={Masahiro Negishi and Hyunsoo Park and Kinga O. Mastej and Aron Walsh},
      year={2025},
      journal={arXiv preprint arXiv:2510.12405}
}

@article{MatterGen2025,
  author  = {Zeni, Claudio and Pinsler, Robert and Z{\"u}gner, Daniel and Fowler, Andrew and Horton, Matthew and Fu, Xiang and Wang, Zilong and Shysheya, Aliaksandra and Crabb{\'e}, Jonathan and Ueda, Shoko and Sordillo, Roberto and Sun, Lixin and Smith, Jake and Nguyen, Bichlien and Schulz, Hannes and Lewis, Sarah and Huang, Chin-Wei and Lu, Ziheng and Zhou, Yichi and Yang, Han and Hao, Hongxia and Li, Jielan and Yang, Chunlei and Li, Wenjie and Tomioka, Ryota and Xie, Tian},
  journal = {Nature},
  title   = {A generative model for inorganic materials design},
  year    = {2025},
  doi     = {10.1038/s41586-025-08628-5},
}

@InProceedings{joshi2025allatomdiffusiontransformersunified,
  title = 	 {All-atom Diffusion Transformers: Unified generative modelling of molecules and materials},
  author =       {Joshi, Chaitanya K. and Fu, Xiang and Liao, Yi-Lun and Gharakhanyan, Vahe and Miller, Benjamin Kurt and Sriram, Anuroop and Ulissi, Zachary Ward},
  booktitle = 	 {Proceedings of the 42nd International Conference on Machine Learning},
  pages = 	 {28393--28417},
  year = 	 {2025},
  editor = 	 {Singh, Aarti and Fazel, Maryam and Hsu, Daniel and Lacoste-Julien, Simon and Berkenkamp, Felix and Maharaj, Tegan and Wagstaff, Kiri and Zhu, Jerry},
  volume = 	 {267},
  series = 	 {Proceedings of Machine Learning Research},
  month = 	 {13--19 Jul},
  publisher =    {PMLR},
  url = 	 {https://proceedings.mlr.press/v267/joshi25a.html}
}

@inproceedings{
xie2022crystal,
title={Crystal Diffusion Variational Autoencoder for Periodic Material Generation},
author={Tian Xie and Xiang Fu and Octavian-Eugen Ganea and Regina Barzilay and Tommi S. Jaakkola},
booktitle={International Conference on Learning Representations},
year={2022},
url={https://openreview.net/forum?id=03RLpj-tc_}
}

@inproceedings{
jiao2024crystalstructurepredictionjoint,
title={Crystal Structure Prediction by Joint Equivariant Diffusion},
author={Rui Jiao and Wenbing Huang and Peijia Lin and Jiaqi Han and Pin Chen and Yutong Lu and Yang Liu},
booktitle={Thirty-seventh Conference on Neural Information Processing Systems},
year={2023},
url={https://openreview.net/forum?id=DNdN26m2Jk}
}

@inproceedings{
jiao2024spacegroupconstrainedcrystal,
title={Space Group Constrained Crystal Generation},
author={Rui Jiao and Wenbing Huang and Yu Liu and Deli Zhao and Yang Liu},
booktitle={The Twelfth International Conference on Learning Representations},
year={2024},
url={https://openreview.net/forum?id=jkvZ7v4OmP}
}

@article{fcd,
author = {Preuer, Kristina and Renz, Philipp and Unterthiner, Thomas and Hochreiter, Sepp and Klambauer, G{\"u}nter},
title = {Fréchet ChemNet Distance: A Metric for Generative Models for Molecules in Drug Discovery},
journal = {Journal of Chemical Information and Modeling},
volume = {58},
number = {9},
pages = {1736-1741},
year = {2018},
doi = {10.1021/acs.jcim.8b00234},
    note ={PMID: 30118593},
URL = { https://doi.org/10.1021/acs.jcim.8b00234
},
eprint = {    https://doi.org/10.1021/acs.jcim.8b00234

}

}

@inproceedings{
duval2025lematgenbench,
title={LeMat-GenBench: Bridging the gap between crystal generation and materials discovery},
author={Alexandre Duval and Siddharth Betala and Samuel P. Gleason and Andy Xu and Georgia Channing and Daniel Levy and Ali Ramlaoui and Cl{\'e}mentine Fourrier and Chaitanya K. Joshi and Nikita Kazeev and S{\'e}kou-Oumar Kaba and F{\'e}lix Therrien and Alex Hern{\'a}ndez-Garc{\'\i}a and Roc{\'\i}o Mercado and N M Anoop Krishnan},
booktitle={AI for Accelerated Materials Design - NeurIPS 2025},
year={2025},
url={https://openreview.net/forum?id=ZfPGcTfDWn}
}

@article{llmcrystalgeneration,
    author = {Hyunsoo Park and Anthony Onwuli and Aron Walsh},
    title = {Exploration of crystal chemical space using text-guided generative artificial intelligence},
    journal = {Nat Commun},
    volume = {16},
    year = {2025},
    doi = {https://doi.org/10.1038/s41467-025-59636-y}
}

@article{materialsproject,
    author = {Jain, Anubhav and Ong, Shyue Ping and Hautier, Geoffroy and Chen, Wei and Richards, William Davidson and Dacek, Stephen and Cholia, Shreyas and Gunter, Dan and Skinner, David and Ceder, Gerbrand and Persson, Kristin A.},
    title = {Commentary: The Materials Project: A materials genome approach to accelerating materials innovation},
    journal = {APL Materials},
    volume = {1},
    number = {1},
    pages = {011002},
    year = {2013},
    month = {07},
    issn = {2166-532X},
    doi = {10.1063/1.4812323},
    url = {https://doi.org/10.1063/1.4812323},
    eprint = {https://pubs.aip.org/aip/apm/article-pdf/doi/10.1063/1.4812323/13163869/011002_1_online.pdf},
}

@misc{disorderdcompounds,
      title={Continued Challenges in High-Throughput Materials Predictions: MatterGen predicts compounds from the training dataset}, 
      author={Mikkel Juelsholt},
      year={2025},
      archivePrefix={ChemRxiv},
      url={https://chemrxiv.org/engage/chemrxiv/article-details/685057d9c1cb1ecda05e9aec}
}

@article{wbmdataset,
    author = {Hai-Chen Wang and Miguel A. L. Marques and Silvana Botti},
    title = {Predicting stable crystalline compounds using chemical similarity},
    journal = {npj Comput Mater},
    volume = {7},
    year = {2021},
    doi = {https://doi.org/10.1038/s41524-020-00481-6}
}

@article{criticofgnome,
    author = {Cheetham, Anthony K. and Seshadri, Ram},
    title = {Artificial Intelligence Driving Materials Discovery? Perspective on the Article: Scaling Deep Learning for Materials Discovery},
    journal = {Chemistry of Materials},
    volume = {36},
    number = {8},
    pages = {3490-3495},
    year = {2024},
    doi = {10.1021/acs.chemmater.4c00643}
}

@article{disordermargraf,
author = {Jakob, Konstantin S. and Walsh, Aron and Reuter, Karsten and Margraf, Johannes T.},
title = {Learning Crystallographic Disorder: Bridging Prediction and Experiment in Materials Discovery},
journal = {Advanced Materials},
volume = {n/a},
number = {n/a},
pages = {e14226},
doi = {https://doi.org/10.1002/adma.202514226},
url = {https://advanced.onlinelibrary.wiley.com/doi/abs/10.1002/adma.202514226},
eprint = {https://advanced.onlinelibrary.wiley.com/doi/pdf/10.1002/adma.202514226}
}

@misc{metni2025generativemodelscrystallinematerials,
      title={Generative models for crystalline materials}, 
      author={Houssam Metni and Laura Ruple and Lauren N. Walters and Luca Torresi and Jonas Teufel and Henrik Schopmans and Jona Östreicher and Yumeng Zhang and Marlen Neubert and Yuri Koide and Kevin Steiner and Paul Link and Lukas Bär and Mariana Petrova and Gerbrand Ceder and Pascal Friederich},
      year={2025},
      eprint={2511.22652},
      archivePrefix={arXiv},
      primaryClass={cond-mat.mtrl-sci},
      url={https://arxiv.org/abs/2511.22652}, 
}

@article{park2025closingsynt,
  title        = {Closing the synthesis gap in computational materials design},
  author       = {Park, H. and Mastej, K. O. and Detrattanawichai, P. and Nduma, R. and Walsh, A.},
  year         = {2025},
  journal      = {ChemRxiv},
  doi          = {10.26434/chemrxiv-2025-sbc0c-v2},
  url          = {https://doi.org/10.26434/chemrxiv-2025-sbc0c-v2},
}

@Article{D4DD00394B,
author ="Amariamir, Sasan and George, Janine and Benner, Philipp",
title  ="SynCoTrain: a dual classifier PU-learning framework for synthesizability prediction",
journal  ="Digital Discovery",
year  ="2025",
volume  ="4",
issue  ="6",
pages  ="1437-1448",
publisher  ="RSC",
doi  ="10.1039/D4DD00394B",
url  ="http://dx.doi.org/10.1039/D4DD00394B",
}

\end{document}